\documentclass[aps,prd,preprint,superscriptaddress,amsmath,amssymb,showpacs]{revtex4-1}
\usepackage{xcolor}
\usepackage{dcolumn}
\usepackage{graphicx}
\usepackage{float}
\usepackage{physics}
\usepackage[colorlinks=true,allcolors=blue]{hyperref}
\begin{document}
	\title{Causality and stability analysis of relativistic spin hydrodynamics: Insights from a nonvanishing spin density background}
	
	\author{Wei Lu}
	\affiliation{College of Mathematics and Physics, China Three Gorges University, Yichang 443002, China}
	
	\author{Yang Zhong}
	\email{zhy@ctgu.edu.cn}
	\affiliation{College of Mathematics and Physics, China Three Gorges University, Yichang 443002, China}
	\affiliation{Center for Astronomy and Space Sciences and Institute of Modern Physics, China Three Gorges University, Yichang 443002, China}
	
	\author{Sheng-Qin Feng}
	\email{fengsq@ctgu.edu.cn}
	\affiliation{College of Mathematics and Physics, China Three Gorges University, Yichang 443002, China}
	\affiliation{Center for Astronomy and Space Sciences and Institute of Modern Physics, China Three Gorges University, Yichang 443002, China}
	\affiliation{Key Laboratory of Quark and Lepton Physics (MOE) and Institute of Particle Physics,\\
		Central China Normal University, Wuhan 430079, China}	

\date{\today}%

	\begin{abstract}
		 We investigate the stability and causality of relativistic spin hydrodynamics in the presence of a nonvanishing spin density background, assuming that the spin chemical potential $\omega^{\mu\nu}$ is of leading order ($\omega^{\mu\nu} \sim \mathcal{O}(1)$) in the gradient expansion and is treated as a finite background in the linear perturbation analysis.
It is found that within the first-order spin hydrodynamic framework, a finite spin density background modifies the dispersion relations, and modes propagating along different directions are controlled by distinct transport coefficients. Certain specific modes only appear in the $x$ direction. However, the modes in the large wave-vector limit exhibit acausal behavior. To address this issue, we subsequently adopt the framework of minimal causal spin hydrodynamics and derive the corresponding stability and causality conditions. The spin density background directly determines whether stability and causality can be satisfied simultaneously. In the small wave-vector limit, the results are similar to those of the first-order theory. In the large wave-vector region, however, significant differences emerge: the distinctions between different directions are no longer merely simple substitutions of transport coefficients, but involve more complex combinations. This indicates that the difference between modes in different directions increases with increasing wave vector.
	\end{abstract}
	
\maketitle
	
\section{Introduction}	
High-energy heavy-ion collision experiments provide a unique opportunity to create extremely hot and dense matter under controlled laboratory conditions, enabling the investigation of the quark-gluon plasma (QGP) that is believed to have filled the early Universe. In such complex many-particle systems, relativistic hydrodynamics has proven remarkably successful in describing macroscopic collective phenomena \cite{PhysRevLett.125.012301,Becattini:2024uha,PhysRevC.98.014910,PhysRevC.101.044611,STAR:2020xbm,PhysRevLett.128.172005}, such as various flow observables, indicating that the produced medium behaves as a nearly perfect fluid over short timescales \cite{Heinz:2013th,Gale:2013da,Shen:2020mgh}.

In recent years, experimental measurements have unveiled a series of intriguing spin-related phenomena, including the significant global polarization of $\Lambda$ hyperons \cite{STAR:2017ckg, Niida:2018hfw,PhysRevC.104.L061901,PhysRevLett.123.132301,PhysRevLett.120.012302,STAR:2020xbm} and the spin alignment of vector mesons \cite{PhysRevLett.125.012301,Singha:2022syo,STAR:2022fan,Schilling:1969um,PhysRevLett.131.042303}. These observations suggest that in noncentral heavy-ion collisions, the QGP acquires substantial total orbital angular momentum, part of which can be converted into spin angular momentum through spin-orbit coupling and subsequently transferred to final-state particles \cite{Liang:2004xn,PhysRevLett.94.102301}. This discovery opens a new avenue for probing the microscopic structure and dynamics of the QGP via spin degrees of freedom. However, conventional relativistic hydrodynamics does not incorporate spin as an intrinsic degree of freedom and is therefore unable to describe the generation, evolution, and transport of spin.

In response to these developments, relativistic spin hydrodynamics has emerged as a natural extension of conventional hydrodynamic theory. In addition to energy-momentum conservation, this framework introduces dynamical variables associated with spin angular momentum density and currents, along with their corresponding conservation laws, aiming to provide a self-consistent macroscopic description of relativistic fluids with spin. In recent years, spin hydrodynamics has advanced rapidly. Various formulations have been developed based on different microscopic or first-principles approaches, such as entropy-current and generalized thermodynamic analyses ~\cite{Hongo:2021ona,Wang:2021ngp,Speranza:2021bxf,Li:2020eon,Hattori:2019lfp,Fukushima:2020ucl,She:2021lhe}, relativistic kinetic theory ~\cite{Florkowski:2018myy,Peng:2021ago,Florkowski:2017ruc,Hidaka:2017auj,Weickgenannt:2019dks,Sheng:2021kfc,Fang:2022ttm,Hu:2021pwh,Weickgenannt:2020aaf}, quantum statistical density operators \cite{Becattini:2007nd,Becattini:2009wh,Becattini:2012pp,Hu:2021lnx,Becattini:2012tc,Becattini:2018duy}, effective field theory~\cite{Montenegro:2017rbu,Montenegro:2017lvf,Montenegro:2018bcf,Montenegro:2020paq,Liu:2020dxg}, holographic duality \cite{Gallegos:2020otk,Garbiso:2020puw}, and equilibrium partition function techniques.

As in conventional hydrodynamics, a physically consistent and numerically reliable theory of spin hydrodynamics must satisfy two fundamental requirements: stability and causality. These conditions are essential for the internal consistency of the theoretical framework and for the robustness of numerical simulations. Following the pioneering work of Hiscock and Lindblom \cite{Hiscock:1983zz,Hiscock:1985zz,Hiscock:1987zz}, subsequent studies on the stability and causality of spin hydrodynamics have continued to advance. It is worth noting that most existing studies in this area perform linear perturbation analyses under the assumption of a vanishing background spin density \cite{Daher:2022wzf,Daher:2024bah,Sarwar:2022yzs,Xie:2023gbo}. However, in realistic physical environments such as the QGP produced in heavy-ion collisions, the large initial orbital angular momentum is expected to induce a nonvanishing background spin density. In such scenarios, the coupling between spin degrees of freedom and hydrodynamic variables -- including fluid velocity and energy density -- may significantly modify the system's dynamical behavior, thereby introducing new and more general challenges to the stability and causality of the theory.

In this work, we investigate the stability and causality of relativistic spin hydrodynamics in the presence of a nonvanishing background spin density, assuming that the spin chemical potential enters at leading order, i.e., $\omega^{\mu\nu} \sim \mathcal{O}(1)$ , and remains finite in the linear perturbation analysis. It is found that within the first-order theory, dissipative modes induced by the finite spin density violate causality, rendering the first-order formulation intrinsically acausal. To resolve this issue, it becomes necessary to consider second-order hydrodynamic theories \cite{Israel:1979wp,10.1098/rspa.1979.0005,Baier:2007ix,Denicol:2012cn}. In this study, a minimal second-order framework that retains only the relaxation-time terms from the M$\ddot{\textrm{u}}$ller-Israel-Stewart (MIS) theory is adopted to analyze stability and causality \cite{Xie:2023gbo,Koide:2006ef}. Our results demonstrate that this second-order formulation successfully eliminates the acausal modes.

This paper is organized as follows. In Sec. \ref{chap:2}, we briefly review relativistic spin hydrodynamics and perform a mode analysis of the first-order theory. A mode analysis of the minimal causal second-order theory is carried out in Sec. \ref{chap:3}. Finally, Sec. \ref{chap:4} is devoted to conclusions and discussions.	
	
Throughout this work, we adopt the metric convention $g_{\mu\nu}=diag\left\{ +,-,-,-\right\}$. The four-velocity satisfies the normalization condition $u^{\mu}u_{\mu}=1$, and the projection tensor orthogonal to the four-velocity is defined as $\Delta^{\mu\nu}=g^{\mu\nu}-u^{\mu}u^{\nu}$.The projection of a four-vector onto the subspace orthogonal to the four-velocity is denoted by $A^{\left\langle \mu\right\rangle }=\Delta^{\mu\nu}A_{\nu}$. For an arbitrary rank-two tensor, its symmetric and antisymmetric parts are denoted by $A_{(s)}^{\mu\nu}=A^{(\mu\nu)}=\frac{1}{2}\left(A^{\mu\nu}+A^{\nu\mu}\right)$ and $A_{(a)}^{\mu\nu}=A^{[\mu\nu]}=\frac{1}{2}\left(A^{\mu\nu}-A^{\nu\mu}\right)$ respectively. The symmetric, traceless, and four-velocity orthogonal part of a rank-two tensor is denoted by $A^{\left\langle \mu\nu\right\rangle }=\Delta_{\alpha\beta}^{\mu\nu}A^{\alpha\beta}=\frac{1}{2}\left(\Delta_{\ \alpha}^{\mu}\Delta_{\ \beta}^{\nu}+\Delta_{\ \beta}^{\mu}\Delta_{\ \alpha}^{\nu}-\frac{2}{3}\Delta^{\mu\nu}\Delta_{\alpha\beta}\right)A^{\alpha\beta}$. The contraction of tensor indices can be denoted by $A^\mu B_\mu \equiv A\cdot B$.

\section{Linear mode analysis of first-order spin hydrodynamics}\label{chap:2}	
 \subsection{Relativistic spin hydrodynamics}

The hydrodynamic equations are built upon the fundamental conservation laws of the system, including the conservation of particle number, energy, momentum, and total angular momentum as
\begin{align}\label{eq:1}
 \partial_{\mu}N^{\mu}=0,\quad\partial_{\mu}T^{\mu\nu}=0,\quad\partial_{\lambda}J^{\lambda\mu\nu}=0,
\end{align}
where $N^{\mu}$ denotes the particle-number current, $T^{\mu\nu}$ is the energy-momentum tensor, and $J^{\lambda\mu\nu}$ is the total angular momentum tensor. The total angular momentum tensor $J^{\lambda\mu\nu}$ can be written as
\begin{align}\label{eq:2}
 J^{\lambda\mu\nu}=x^{\mu}T^{\lambda\nu}-x^{\nu}T^{\lambda\mu}+\Sigma^{\lambda\mu\nu},
\end{align}
where the first two terms $x^{\mu}T^{\lambda\nu}-x^{\nu}T^{\lambda\mu}$ represent the orbital angular momentum $L^{\lambda\mu\nu}$ , while $\Sigma^{\lambda\mu\nu}$ denotes the spin angular momentum, which satisfies $\Sigma^{\lambda\mu\nu}=-\Sigma^{\lambda\nu\mu}$. Combining Eqs. \eqref{eq:1} and \eqref{eq:2}, one obtains
\begin{align}\label{eq:3}
 \partial_{\lambda}\Sigma^{\lambda\mu\nu}=-2T^{[\mu\nu]}.
\end{align}
The right-hand side of the above equation originates from the orbital angular momentum part in the total angular momentum expression of Eq. \eqref{eq:2}, indicating that spin and orbital angular momenta can be converted into each other. Moreover, the antisymmetric part of the energy-momentum tensor acts as a source or sink for the spin current \cite{Hattori:2019lfp,Xie:2023gbo}.

Herein, we further elaborate on the issue regarding the choice of forms for the energy-momentum tensor and the angular momentum tensor.
In quantum field theory, according to Noether's theorem, translational symmetry and Lorentz symmetry -- both continuous symmetries -- each correspond to a conserved charge, namely the canonical energy-momentum tensor and the canonical angular momentum tensor (the subscript "can" is used to denote the canonical form):
\begin{align}
 T^{\mu\nu}_{\text{can}}=&\frac{\partial\mathcal{L}}{\partial\big(\partial_{\mu}\Phi_{a}\big)}\partial^{\nu}\Phi_{a}-g^{\mu\nu}\mathcal{L}, \\
 J^{\lambda\mu\nu}_{\text{can}}
 =&x^{\mu}T^{\lambda\nu}_{\text{can}}-x^{\nu}T^{\lambda\mu}_{\text{can}}+\Sigma^{\lambda\mu\nu}_{\text{can}},
\end{align}
where $\mathcal{L}$ is the Lagrangian density and $\Phi_{a}$ is the field variable, in terms of which the spin tensor can then be explicitly expressed as
\begin{align}
 \Sigma^{\lambda\mu\nu}_{\text{can}}
 =&-\mathrm{i}\frac{\partial\mathcal{L}}{\partial\big(\partial_{\lambda}\Phi_{a}\big)}\big(I^{\mu\nu}\big)_{ab}\Phi_{b},
\end{align}
where $\big(I^{\mu\nu}\big)_{ab}$ is the generator of the representation of the Lorentz group to which $\Phi_{a}$ belongs. However, the energy-momentum tensor and the angular momentum tensor are not uniquely determined. By means of the so-called pseudogauge transformation, distinct forms can be derived without altering the conservation properties of the energy-momentum tensor and the total angular momentum tensor \cite{Fukushima:2020ucl,Florkowski:2018fap,Hehl:1976vr,Speranza:2020ilk}. The transformations read as follows
\begin{align}
 \widetilde{T}^{\mu\nu}=&T_{\text{can}}^{\mu\nu}+\frac{1}{2}\partial_{\lambda}\Big(\varPhi^{\lambda\mu\nu}-\varPhi^{\mu\lambda\nu}+\varPhi^{\nu\mu\lambda}\Big), \\
 \widetilde{\Sigma}^{\lambda\mu\nu}=&\Sigma_{\text{can}}^{\lambda\mu\nu}-\varPhi^{\lambda\mu\nu},
\end{align}
where $\varPhi^{\lambda\mu\nu}$ is a rank-three tensor field, which is called a superpotential. We denote the new energy-momentum tensor and spin tensor by $\widetilde{T}^{\mu\nu}$ and $\widetilde{\Sigma}^{\lambda\mu\nu}$, respectively. For example, if one chooses $\varPhi^{\lambda\mu\nu}=\Sigma_{\text{can}}^{\lambda\mu\nu}$, the spin tensor vanishes, the new energy-momentum tensor takes a symmetric form, and the total angular momentum tensor no longer contains an explicit spin component. Their respective expressions are given by
\begin{align}
 \mathcal{T}^{\mu\nu}_{\text{Bel}}=&T_{\text{can}}^{\mu\nu}+\frac{1}{2}\partial_{\lambda}\Big(\Sigma_{\text{can}}^{\lambda\mu\nu}-\Sigma_{\text{can}}^{\mu\lambda\nu}+\Sigma_{\text{can}}^{\nu\mu\lambda}\Big),\\
 \mathcal{J}^{\lambda\mu\nu}_{\text{Bel}}=&x^{\mu}\mathcal{T}^{\lambda\nu}_{\text{Bel}}-x^{\nu}\mathcal{T}^{\lambda\mu}_{\text{Bel}},\\
 \Sigma^{\lambda\mu\nu}_{\text{Bel}}=&0,
\end{align}
where $\mathcal{T}^{\mu\nu}_{\text{Bel}}$, $\mathcal{J}^{\lambda\mu\nu}_{\text{Bel}}$, and $\Sigma^{\lambda\mu\nu}_{\text{Bel}}$ are the Belinfante forms of the energy-momentum tensor, the total angular momentum tensor, and the spin angular momentum tensor, respectively. 

More generally, the canonical form of the spin angular momentum tensor is antisymmetric only in its last two indices. However, a totally antisymmetric form with respect to all indices can be constructed by modifying the Lagrangian density~\cite{Xie:2023gbo}. The two equivalent Lagrangian densities are given as 
\begin{align}
 \mathcal{L}_{1}=&\bar{\Phi}\left(i\gamma\cdot\partial-m\right)\Phi,\\\mathcal{L}_{2}=&\frac{1}{2}\bar{\Phi}i\gamma\cdot\overleftrightarrow{\partial}\Phi-m\bar{\Phi}\Phi,
\end{align}
where $\gamma^{\mu}$ is the gamma matrix, and $\overleftrightarrow{\partial}_{\mu}\equiv\overrightarrow{\partial}_{\mu}-\overleftarrow{\partial}_{\mu}$. Two sets of energy-momentum and spin tensors can be obtained accordingly
\begin{align}
T_{1}^{\mu\nu}=\bar{\Phi}i\gamma^{\mu}\partial^{\nu}\Phi,\quad \Sigma_{1}^{\lambda\mu\nu}=\frac{1}{4}\bar{\Phi}i\gamma^{\lambda}\left[\gamma^{\mu},\gamma^{\nu}\right]\Phi,
\end{align}
\begin{align}
T_{2}^{\mu\nu}=\frac{i}{2}\bar{\Phi}\gamma^{\mu}\overleftrightarrow{\partial^{\nu}}\Phi,\quad\Sigma_{2}^{\lambda\mu\nu}=\frac{1}{8}\bar{\Phi}i\left\{ \gamma^{\lambda},\left[\gamma^{\mu},\gamma^{\nu}\right]\right\} \Phi,
\end{align}
where $\Sigma_{1}^{\lambda\mu\nu}$ corresponds to the form that is antisymmetric in its last two indices, while $\Sigma_{2}^{\lambda\mu\nu}$ corresponds to the form fully antisymmetric in all indices. In the theory of spin hydrodynamics, different choices of tensor forms yield distinct results, and the question of which form conforms to objective physical reality remains an open issue. In this work, we employ the canonical forms of the energy-momentum tensor and the angular momentum tensor. The stability and causality results derived subsequently via linear perturbation analysis may not be applicable to other forms. The conditions corresponding to different tensor forms should be analyzed independently.

The particle-number current, energy-momentum tensor, and angular momentum tensor can be decomposed with respect to the four-velocity as~\cite{Hongo:2021ona,Li:2020eon,Hattori:2019lfp,Fukushima:2020ucl,She:2021lhe,Dey:2024cwo,Biswas:2023qsw}
\begin{gather}
 N^{\mu}=N_{(0)}^{\mu}+N_{(1)}^{\mu}, \\
 N_{(0)}^{\mu}=nu^{\mu},\quad N_{(1)}^{\mu}=n^{\mu}, \\
 T^{\mu\nu}=T_{(0)}^{\mu\nu}+T_{(1)}^{\mu\nu}, \\
 T_{(0)}^{\mu\nu}=eu^{\mu}u^{\nu}-p\Delta^{\mu\nu}, \\
 T_{(1)}^{\mu\nu}=2h^{(\mu}u^{\nu)}-\Pi\Delta^{\mu\nu}+\pi^{\mu\nu}+2q^{[\mu}u^{\nu]}+\phi^{\mu\nu}, \\
 \Sigma^{\lambda\mu\nu}=\Sigma_{(0)}^{\lambda\mu\nu}+\Sigma_{(1)}^{\lambda\mu\nu}, \\
 \Sigma_{(0)}^{\lambda\mu\nu}=u^{\lambda}S^{\mu\nu}, \\
 \Sigma_{(1)}^{\lambda\mu\nu}=2\Delta^{\lambda[\mu}u^{\nu]}\Sigma+2\Sigma_{(s)}^{\left\langle \lambda[\mu\right\rangle }u^{\nu]}+2\Sigma_{(a)}^{[\lambda[\mu]}u^{\nu]}+\Sigma^{\left\langle \lambda\right\rangle \left\langle \mu\right\rangle \left\langle \nu\right\rangle },
\end{gather}
where $N_{(0)}^{\mu}$, $T_{(0)}^{\mu\nu}$, and $\Sigma_{(0)}^{\lambda\mu\nu}$ denote the zeroth-order contributions of the corresponding physical quantities, respectively, while $N_{(1)}^{\mu}$, $T_{(1)}^{\mu\nu}$, and $\Sigma_{(1)}^{\lambda\mu\nu}$ represent the corresponding first-order derivative corrections, respectively. $e$, $p$, $n$, and $S^{\mu\nu}$ are the energy density, pressure, particle number density, and spin density, respectively, whose orders in the gradient expansion are given by $\mathcal{O}\left(1\right)$. $n^{\mu}$, $h^{\mu}$, $\Pi$, and $\pi^{\mu\nu}$ denote the particle diffusion current, heat flow, bulk viscous pressure, and shear stress tensor, respectively. $q^{\mu}$ and $\phi^{\mu\nu}$ correspond to the antisymmetric part of the energy-momentum tensor and is associated with spin effects, while $\Sigma$, $\Sigma_{(s)}^{\left\langle \mu\alpha\right\rangle }$, $\Sigma_{(a)}^{[\mu\alpha]}$, and $\Sigma^{\left\langle \mu\right\rangle \left\langle \alpha\right\rangle \left\langle \beta\right\rangle }$ represent spin-related dissipative currents of the corresponding physical quantities, respectively. All dissipative currents introduced above satisfy the following constraints: $n^{\mu}u_{\mu}=0$, $h^{\mu}u_{\mu}=0$, $\pi^{\mu\nu}u_{\mu}=\pi^{\mu\nu}u_{\nu}=0$, $q^{\mu}u_{\mu}=0$, $\phi^{\mu\nu}u_{\mu}=\phi^{\mu\nu}u_{\nu}=0$, $\Sigma_{(s)}^{\left\langle \mu\alpha\right\rangle }u_{\mu}=\Sigma_{(s)}^{\left\langle \mu\alpha\right\rangle }u_{\alpha}=0$, $\Sigma_{(a)}^{[\mu\alpha]}u_{\mu}=\Sigma_{(a)}^{[\mu\alpha]}u_{\alpha}=0$, and $u_{\mu}\Sigma^{\left\langle \mu\right\rangle \left\langle \alpha\right\rangle \left\langle \beta\right\rangle }=u_{\alpha}\Sigma^{\left\langle \mu\right\rangle \left\langle \alpha\right\rangle \left\langle \beta\right\rangle }=u_{\beta}\Sigma^{\left\langle \mu\right\rangle \left\langle \alpha\right\rangle \left\langle \beta\right\rangle }=0$, namely, all first-order dissipative currents are orthogonal to the four-velocity.

The particle-number current $N^{\mu}$ possesses four independent components. The particle number density $n$ is a scalar with one degree of freedom, while the dissipative particle current $n^{\mu}$ is a four-vector orthogonal to the four-velocity and therefore has three independent components. Consequently, the degrees of freedom are matched in the decomposition of the particle-number current. The spin angular momentum tensor $\Sigma^{\lambda\mu\nu}$ is a rank-three tensor antisymmetric in its last two indices, carrying a total of 24 degrees of freedom. The spin density $S^{\mu\nu}$ is an antisymmetric tensor with six independent components. The scalar $\Sigma$ has one degree of freedom. The tensor $\Sigma_{(s)}^{\left\langle \mu\alpha\right\rangle }$ is symmetric, traceless, and orthogonal to the four-velocity, and thus contains five independent components. The tensor $\Sigma_{(a)}^{[\mu\alpha]}$ is antisymmetric and orthogonal to the four-velocity, with three degrees of freedom. The tensor $\Sigma^{\left\langle \mu\right\rangle \left\langle \alpha\right\rangle \left\langle \beta\right\rangle }$ is antisymmetric in its last two indices and orthogonal to the four-velocity in all indices, possessing nine independent components. Therefore, the degrees of freedom are consistently matched in the decomposition of the spin tensor. For the energy-momentum tensor $T^{\mu\nu}$, there are in total 16 independent components. Similarly, the scalar $e$, $\Pi$ each carry one degree of freedom, each of the vectors $h^\mu$, $q^\mu$ has three degrees of freedom, and the four-velocity itself has only three independent components due to the normalization condition. The tensor $\pi^{\mu\nu}$ is symmetric, traceless, and orthogonal to the four-velocity, contributing five degrees of freedom, while the antisymmetric tensor $\phi^{\mu\nu}$ has three degrees of freedom. Altogether, this yields 19 degrees of freedom. To eliminate the redundant three degrees of freedom and achieve consistency with the energy-momentum tensor, one must fix the definition of the fluid velocity. There exist several possible choices for the fluid frame, among which the two most common are the Eckart frame and the Landau frame. In this work, we adopt the Landau frame, in which $h^\mu=0$ .

With the inclusion of spin as an additional degree of freedom in hydrodynamics, the thermodynamic relations are correspondingly modified as \cite{Hongo:2021ona,Wang:2021ngp,Speranza:2021bxf,Li:2020eon,Hattori:2019lfp,Fukushima:2020ucl,She:2021lhe,Dey:2024cwo,Biswas:2023qsw}
\begin{align}
 e+p=Ts+\mu n+\omega_{\alpha\beta}S^{\alpha\beta}, \label{eq:103} \\
 {\rm d}e=T{\rm d}s+\mu{\rm d}n+\omega_{\alpha\beta}{\rm d}S^{\alpha\beta}, \label{eq:104} \\
 {\rm d}p=s{\rm d}T+n{\rm d}\mu+S^{\alpha\beta}{\rm d}\omega_{\alpha\beta}, \label{eq:105}
\end{align}
where $T$, $s$, $\mu$ and $\omega_{\alpha\beta}$ denote the temperature, entropy density, chemical potential, and spin chemical potential, respectively. $\omega_{\alpha\beta}$ and $\mu$ play analogous roles, and are both introduced as hydrodynamic variables, serving as the conjugate quantities to $S^{\alpha\beta}$ and $n$, respectively. Although the particle number density, which is conjugate to the chemical potential, is conserved, the spin angular momentum conjugating to the spin chemical potential is not necessarily conserved. As follows from Eq. \eqref{eq:3}, the antisymmetric part of the energy-momentum tensor can induce spin-orbit conversion, in which case only the total angular momentum remains conserved.

Within the framework of the quantum statistical density operator~\cite{Becattini:2012tc,Florkowski:2018ahw}, it can be shown that the spin chemical potential $\omega^{\mu\nu}$ is completely determined by the thermal vorticity $\varpi_{\mu\nu}\equiv-\left(\partial_{\mu}\beta_{\nu}-\partial_{\nu}\beta_{\mu}\right)/2$~ in global equilibrium. In this approach, the most general condition for global equilibrium takes the form
\begin{align}
 \partial_{(\mu}\beta_{\nu)}=0,\quad\beta_\mu=b_\mu+\varpi_{\mu\nu}x^\nu,  \label{eq:4}   \\
 \varpi_{\mu\nu}=-\partial_{[\mu}\beta_{\nu]}=constant,         \label{eq:5}
\end{align}
where $\beta_\mu=\beta u_\mu$, $\beta=1/T$, and $b_\mu$ is a constant four-vector. From Eqs. \eqref{eq:4} and \eqref{eq:5} one finds $\varpi_{\mu\nu} \sim \mathcal{O}(\partial)$. In global equilibrium the spin chemical potential is completely determined by the thermal vorticity, so it is natural to take $\omega^{\mu\nu} \sim \mathcal{O}(\partial)$. This chain of reasoning rests on the energy-momentum tensor possessing an antisymmetric part. If instead the energy-momentum tensor is taken to be symmetric, spin angular momentum becomes an independently conserved quantity [Eq. \eqref{eq:3}], and the link between the spin chemical potential and the thermal vorticity in global equilibrium is severed \cite{Dey:2024cwo}. Consequently, $\omega^{\mu\nu}$ is no longer forced to be 
$\mathcal{O}(\partial)$ and can consistently be counted as $\mathcal{O}(1)$.

When we study linear stability and causality with a nonvanishing background spin density, the background state necessarily carries a finite spin chemical potential. In the antisymmetric framework this would also require a finite thermal vorticity to be compatible with the quantum statistical density operator— i.e., a rigidly rotating background—which makes the analysis substantially more involved. To keep the background static and simplify the calculations, we therefore adopt a symmetric energy-momentum tensor and the corresponding $\mathcal{O}(1)$ counting for $\omega^{\mu\nu}$. In this framework the spin current is independently conserved and the spin chemical potential is not tied to the thermal vorticity, permitting a static background with a finite spin density.

To explicitly distinguish our framework~\cite{She:2021lhe,Dey:2024cwo} from other common conventions in the literature\cite{Hattori:2019lfp,Daher:2024bah,Biswas:2023qsw,Xie:2023gbo}, we note that spin hydrodynamic theories generally fall into two categories based on their power counting schemes. In the first category, often derived from quantum statistical operators in global equilibrium, the spin chemical potential is strictly bound to the thermal vorticity ($\omega^{\mu\nu} \sim \varpi^{\mu\nu} \sim \mathcal{O}(\partial)$), which requires an asymmetric energy-momentum tensor. In contrast, the second category—adopted in this work—utilizes a symmetric energy-momentum tensor, which decouples $\omega^{\mu\nu}$ from the thermal vorticity in equilibrium. This allows $\omega^{\mu\nu}$ to be counted as $\mathcal{O}(1)$. The $\mathcal{O}(1)$ power counting scheme provides a distinct advantage for linear stability analysis: it enables us to investigate a homogeneous static background with a finite spin density, avoiding the spatial dependencies and severe mathematical complications introduced by a rigidly rotating background inherent to the $\mathcal{O}(\partial)$ scheme.

From the thermodynamic relations given in Eqs. \eqref{eq:103}-\eqref{eq:105}, the expression for the covariant entropy-current can be derived as
\begin{align}
 s^{\mu}=&\beta u_{\nu}T_{(0)}^{\mu\nu}+\beta^{\mu}p-\alpha N_{(0)}^{\mu}-\beta\omega_{\alpha\beta}\Sigma_{(0)}^{\mu\alpha\beta}+\beta u_{\nu}T_{(1)}^{\mu\nu}-\alpha N_{(1)}^{\mu}-\beta\omega_{\alpha\beta}\Sigma_{(1)}^{\mu\alpha\beta} \notag \\
 =&s_{(0)}^{\mu}+\beta u_{\nu}T_{(1)}^{\mu\nu}-\alpha N_{(1)}^{\mu}-\beta\omega_{\alpha\beta}\Sigma_{(1)}^{\mu\alpha\beta},
\end{align}
where $s^{\mu}$ is the entropy density current with $\alpha=\mu/T$, and $s_{(0)}^{\mu}$ is the zeroth-order entropy density current; $s_{(0)}^{\mu} \equiv su^\mu \equiv \beta u_{\nu}T_{(0)}^{\mu\nu}+\beta^{\mu}p-\alpha N_{(0)}^{\mu}-\beta\omega_{\alpha\beta}\Sigma_{(0)}^{\mu\alpha\beta}$.

The zeroth-order entropy production can then be written as
\begin{align}
 \partial_{\mu}s_{(0)}^{\mu}=2\beta\omega_{\alpha\beta}T^{[\alpha\beta]}_{(1)},
\end{align}
when the spin chemical potential is taken to be $\mathcal{O}(1)$, the requirement of vanishing entropy production in the ideal fluid limit implies that the antisymmetric part of the energy-momentum tensor must vanish. This result is consistent with the conclusion obtained from the quantum statistical density operator framework~\cite{Becattini:2012tc,Florkowski:2018ahw,Becattini:2015nva,Dey:2024cwo}.

As derived in Refs. \cite{Hattori:2019lfp,Fukushima:2020ucl,She:2021lhe,Dey:2024cwo}, the entropy production rate can be obtained as
\begin{align}
  \partial_{\mu}s^{\mu}
    =&n^{\mu}\partial_{\mu}\alpha+h^{\mu}\Big(\partial_{\mu}\beta+\beta u^{\nu}\partial_{\nu}u_{\mu}\Big)-\Pi\Big(\beta\Delta^{\mu\nu}\partial_{\mu}u_{\nu}\Big)+\pi^{\mu\nu}\Big(\beta\partial_{\mu}u_{\nu}\Big)
    \notag \\
     &-\Sigma\left[2\Delta^{\mu\alpha}u^{\beta}\partial_{\mu}\left(\beta\omega_{\alpha\beta}\right)\right]-\Sigma_{(s)}^{\left\langle \mu\alpha\right\rangle }\left[2u^{\beta}\partial_{\mu}\left(\beta\omega_{\alpha\beta}\right)\right]
     \notag \\
     &-\Sigma_{(a)}^{[\mu\alpha]}\left[2u^{\beta}\partial_{\mu}\left(\beta\omega_{\alpha\beta}\right)\right]-\Sigma^{\left\langle \mu\right\rangle \left\langle \alpha\right\rangle \left\langle \beta\right\rangle }\left[\partial_{\mu}\left(\beta\omega_{\alpha\beta}\right)\right],
\end{align}
requiring consistency with the second law of thermodynamics, i.e. $\partial_{\mu}s^{\mu}\geq0$, one can then derive the first-order constitutive relations \cite{She:2021lhe,Dey:2024cwo} as
\begin{gather}
 n^{\mu}=\kappa_{n}\Delta^{\mu\nu}\left(\partial_{\nu}\alpha\right), \label{eq:6} \\
 h^{\mu}=\kappa\Delta^{\mu\nu}\left(\beta\partial_{\nu}T-u^{\alpha}\partial_{\alpha}u_{\nu}\right), \\
 \Pi=-\zeta\partial_{\mu}u^{\mu}, \\
 \pi^{\mu\nu}=2\eta\partial^{\langle\mu}u^{\nu\rangle}, \\
 \Sigma=-\chi_{1}u^{\beta}\Delta^{\mu\alpha}\partial_{\mu}\left(\beta\omega_{\alpha\beta}\right), \\
 \Sigma_{(s)}^{\left\langle \mu\alpha\right\rangle }=-\chi_{2}u_{\beta}\left(\Delta_{\ \gamma}^{\mu}\Delta_{\ \sigma}^{\alpha}+\Delta_{\ \sigma}^{\mu}\Delta_{\ \gamma}^{\alpha}-\frac{2}{3}\Delta^{\mu\alpha}\Delta_{\gamma\sigma}\right)\partial^{\gamma}\left(\beta\omega^{\sigma\beta}\right), \\
 \Sigma_{(a)}^{[\mu\alpha]}=-\chi_{3}u^{\beta}\Delta^{\mu\zeta}\Delta^{\alpha\lambda}\partial_{[\zeta}\left(\beta\omega_{\lambda]\beta}\right), \\
 \Sigma^{\left\langle \mu\right\rangle \left\langle \alpha\right\rangle \left\langle \beta\right\rangle }=\chi_{4}\Delta^{\mu\rho}\Delta^{\alpha\eta}\Delta^{\beta\kappa}\partial_{\rho}\left(\beta\omega_{\eta\kappa}\right),\label{eq:7}
\end{gather}
where $\kappa$, $\zeta$, and $\eta$ are the heat conductivity, shear and bulk viscous coefficients, respectively. The $\kappa_{n}$, $\chi_{1}$, $\chi_{2}$, $\chi_{3}$, and $\chi_{4}$ are new transport coefficients associated with the spin dissipative currents. The second law of thermodynamics requires all transport coefficients to be positive, i.e.
\begin{align}\label{eq:26}
  \kappa_{n},\kappa,\zeta,\eta,\chi_{1},\chi_{2},\chi_{3},\chi_{4}>0.
\end{align}

The conservation equation Eq. \eqref{eq:1}, together with the constitutive relations \eqref{eq:6}-\eqref{eq:7}, forms a closed set of equations. For simplicity, the $N^\mu$ will be neglected in the following analysis.

 \subsection{Linear mode analysis at first-order}
In this subsection, the stability and causality of first-order spin hydrodynamics are analyzed in the presence of a nonvanishing spin density background. We introduce linear perturbations  around the equilibrium configuration of the hydrodynamic variables $X$, and assume that all perturbations take the plane-wave form $\delta X =\delta\tilde{X}e^{i\omega t-i \vec{k}\cdot \vec{x}}$. Substituting these ans$\ddot{\textrm{a}}$tze into the hydrodynamic equations, one obtains the corresponding dispersion relations $\omega=\omega(k)$. The stability requires
\begin{align}\label{eq:8}
  \mathrm{Im}\omega\left(k\right)>0,
\end{align}
and causality requires
\begin{align}\label{eq:9}
  \underset{k\rightarrow\infty}{\lim}\left|\mathrm{Re}\frac{\omega}{k}\right|\leq1\quad \textrm{or} \quad\underset{k\rightarrow\infty}{\lim}\left|\mathrm{Re}\frac{\partial\omega}{\partial k}\right|\leq1,
\end{align}
and
\begin{align}\label{eq:10}
 \underset{k\rightarrow\infty}{\lim}\left|\frac{\omega}{k}\right| \quad\text{is bounded}.
\end{align}

As pointed out in Refs. \cite{Daher:2024bah,Xie:2023gbo}, Eq. \eqref{eq:9} alone is not sufficient to ensure causality, and Eq. \eqref{eq:10} must be imposed as an additional constraint. Moreover, Eqs. \eqref{eq:9} and \eqref{eq:10} provide only necessary, but not sufficient conditions for causality. Further discussions on stability and causality can be found in Refs. \cite{Wang:2023csj,Gavassino:2021kjm}.

In the existing analyses of spin fluid dynamics stability and causality, the background state is typically set as a static state with zero spin density \cite{Daher:2022wzf,Daher:2024bah,Sarwar:2022yzs,Xie:2023gbo,Koide:2006ef}. A linear perturbation analysis on a static background with nonvanishing spin density is utilized in this work. To simplify the calculations, we assume the background spin density is nonzero only along the $z$ direction, i.e. $S_{(0)}^{0z}$ and $S_{(0)}^{xy}$. The independent hydrodynamic variables are then perturbed as
\begin{align}
 e=e_{(0)}+\delta e,\quad u^\mu=u^\mu_{(0)}+\delta u^\mu, \label{eq:11} \\
 S^{\mu\nu}=\begin{cases}
 S_{(0)}^{0z}+\delta S^{0z} & \mu=0,\nu=z\\
 S_{(0)}^{xy}+\delta S^{xy} & \mu=x,\nu=y\\
 0+\delta S^{\mu\nu} & other
 \end{cases}, \label{eq:12}
\end{align}
where $u_{(0)}^{\mu}=\left(1,\vec{0}\right)$, $\delta u^{\mu}=\left(0,\delta v^{i}\right)$, and $v^{i}$ is the fluid three-velocity.

For the equation of state, the following perturbation relations and parameters are introduced as	
\begin{gather}
\delta p=\chi_{c}\delta e+\mathcal{A}_{0z}\delta S^{0z}+\mathcal{B}_{xy}\delta S^{xy}, \label{eq:13} \\
\delta\beta=\chi_{T}\delta e+\mathcal{C}_{0z}\delta S^{0z}+\mathcal{D}_{xy}\delta S^{xy}, \label{eq:14}\\
\delta\omega^{0i}=\chi_{b}\delta S^{0i},\quad \delta\omega^{ij}=\chi_{s}\delta S^{ij}. \label{eq:15}
\end{gather}

The coefficients introduced in Eqs. \eqref{eq:13}–\eqref{eq:14} have a clear thermodynamic interpretation. In the present framework, the local equilibrium state is characterized by the energy density and the independent spin density components, such that the equation of state can be written as
$p=p(e,S^{0z},S^{xy})$, $\beta=\beta(e,S^{0z},S^{xy})$.
Linearizing these relations around the homogeneous background yields Eqs. \eqref{eq:13}-\eqref{eq:14}, and one obtains the coefficients corresponding to generalized thermodynamic susceptibilities as  
\begin{align}
\chi_c \equiv \left(\frac{\partial p}{\partial e}\right)_{S^{0z},S^{xy}}, \quad \chi_T\equiv \left(\frac{\partial \beta}{\partial e}\right)_{S^{0z},S^{xy}},
\end{align}
which measure the response of the pressure and inverse temperature to energy-density perturbations under fixed spin densities. The remaining coefficients, 
\begin{align}
\mathcal{A}_{0z} \equiv \left(\frac{\partial p}{\partial S^{0z}}\right)_{e,S^{xy}}, \;
\mathcal{B}_{xy} \equiv \left(\frac{\partial p}{\partial S^{xy}}\right)_{e,S^{0z}}, \\
\mathcal{C}_{0z} \equiv \left(\frac{\partial \beta}{\partial S^{0z}}\right)_{e,S^{xy}}, \;
\mathcal{D}_{xy} \equiv \left(\frac{\partial \beta}{\partial S^{xy}}\right)_{e,S^{0z}}, 
\end{align}
describe the coupling between the hydrodynamic and spin sectors and quantify how spin density fluctuations modify the local thermodynamic state.

Coefficients $\chi_b$ and $\chi_s$ are two constant equation-of-state parameters representing the relationship between the spin density tensor and spin chemical potential, corresponding to its "electriclike" and "magneticlike" components. Existing stability analyses indicate that these two parameters have opposite signs \cite{Daher:2022wzf,Xie:2023gbo}
\begin{align}
 \chi_b<0, \quad \chi_s>0. \label{eq:16}
\end{align}

In previous studies, a linear relation between $S^{\mu\nu}$ and $\omega^{\mu\nu}$~ has often been assumed for simplicity. However, such a linear relation can be shown to be invalid under the condition $\chi_b<0$ and $\chi_s>0$. For a detailed investigation of the equation of state relating the spin density and the spin chemical potential, see Ref. \cite{Daher:2024ixz}.

For convenience in writing and calculations, the following definitions of the relevant parameters that will be used subsequently are provided as
\begin{align}
\begin{array}{cccc}
\mathcal{A}=\mathcal{A}_{0z}S_{(0)}^{0z}, & \mathcal{B}=\mathcal{B}_{xy}S_{(0)}^{xy}, & \mathcal{C}=\mathcal{C}_{0z}S_{(0)}^{0z}, & \mathcal{D}=\mathcal{D}_{xy}S_{(0)}^{xy},
\end{array}
\end{align}
\begin{align}
\begin{array}{cccc}
\tilde{\chi}_{1}=\chi_{1}\chi_{b}, & \tilde{\chi}_{2}=\chi_{2}\chi_{b}, & \tilde{\chi}_{3}=\chi_{3}\chi_{b}, & \tilde{\chi}_{4}=\chi_{4}\chi_{s},
\end{array}
\end{align}
\begin{align}
\begin{array}{cc}
\chi_{12}=\chi_{1}+\frac{4}{3}\chi_{2}, & \chi_{23}=\chi_{2}+\chi_{3},
\end{array}
\end{align}
\begin{align}
\begin{array}{cc}
\tilde{\chi}_{12}=\left(\chi_{1}+\frac{4}{3}\chi_{2}\right)\chi_{b}, & \tilde{\chi}_{23}=\left(\chi_{2}+\chi_{3}\right)\chi_{b},
\end{array}
\end{align}
\begin{align}
\begin{array}{ccccc}
\theta=\frac{1}{e_{(0)}+p_{(0)}}, & \mathcal{C}_{T} & =\mathcal{C}+\beta_{(0)}, & \mathcal{D}_{T} & =\mathcal{D}+\beta_{(0)}.
\end{array}
\end{align}

By substituting the perturbations from Eqs. \eqref{eq:11} and \eqref{eq:12} into the spin hydrodynamics Eqs. \eqref{eq:1} and \eqref{eq:6}-\eqref{eq:7}, one obtains the following linearized equations as
\begin{align}
 0&=\partial_{0}\left(\delta e\right)+\partial_{i}\left(\delta\vartheta^{i}\right), \\
 0&=-\chi_{c}\partial^{j}\left(\delta e\right)+\left(\gamma_{\parallel}-\gamma_{\perp}\right)\partial^{j}\partial_{i}\left(\delta\vartheta^{i}\right)+\left(\partial_{0}+\gamma_{\perp}\partial_{i}\partial^{i}\right)\left(\delta\vartheta^{j}\right)-\mathcal{A}_{0z}\partial^{j}\left(\delta S^{0z}\right)-\mathcal{B}_{xy}\partial^{j}\left(\delta S^{xy}\right), \\
 0&=\partial_{0}\left(\delta S^{0i}\right)+S_{(0)}^{0i}\partial_{j}\left(\delta u^{j}\right)-\tilde{\chi}_{1}\chi_{T}S_{(0)0a}\partial^{i}\partial^{a}\delta e-\tilde{\chi}_{1}\mathcal{C}_{0z}S_{(0)0a}\partial^{i}\partial^{a}\delta S^{0z} \notag \\
  &\quad -\tilde{\chi}_{1}\mathcal{D}_{xy}S_{(0)0a}\partial^{i}\partial^{a}\delta S^{xy}-\tilde{\chi}_{1}\beta_{(0)}\partial^{i}\partial_{a}\delta S^{0a}-\tilde{\chi}_{2}\beta_{(0)}\partial_{j}\partial^{i}\delta S^{0j} \notag \\
  &\quad +\tilde{\chi}_{2}\beta_{(0)}\partial_{j}\partial_{j}\delta S^{0i}+\frac{2}{3}\tilde{\chi}_{2}\beta_{(0)}\partial^{i}\partial_{m}\delta S^{0m}-\tilde{\chi}_{2}\chi_{T}S_{(0)}^{0j}\partial_{j}\partial^{i}\delta e \notag \\
  &\quad -\tilde{\chi}_{2}S_{(0)}^{0j}\mathcal{C}_{0z}\partial_{j}\partial^{i}\delta S^{0z}-\tilde{\chi}_{2}S_{(0)}^{0j}\mathcal{D}_{xy}\partial_{j}\partial^{i}\delta S^{xy}+\tilde{\chi}_{2}S_{(0)}^{0i}\chi_{T}\partial_{j}\partial_{j}\delta e \notag \\
  &\quad +\tilde{\chi}_{2}S_{(0)}^{0i}\mathcal{C}_{0z}\partial_{j}\partial_{j}\delta S^{0z}+\tilde{\chi}_{2}S_{(0)}^{0i}\mathcal{D}_{xy}\partial_{j}\partial_{j}\delta S^{xy}+\frac{2}{3}\tilde{\chi}_{2}S_{(0)}^{0m}\chi_{T}\partial^{i}\partial_{m}\delta e \notag \\
  &\quad +\frac{2}{3}\tilde{\chi}_{2}S_{(0)}^{0m}\mathcal{C}_{0z}\partial^{i}\partial_{m}\delta S^{0z}+\frac{2}{3}\tilde{\chi}_{2}S_{(0)}^{0m}\mathcal{D}_{xy}\partial^{i}\partial_{m}\delta S^{xy} \notag \\
  &\quad +\tilde{\chi}_{3}\beta_{(0)}\partial_{j}\partial^{i}\delta S^{0j}+\tilde{\chi}_{3}\beta_{(0)}\partial_{j}\partial_{j}\delta S^{0i}+\tilde{\chi}_{3}S_{(0)}^{0j}\chi_{T}\partial_{j}\partial^{i}\delta e \notag \\
  &\quad +\tilde{\chi}_{3}S_{(0)}^{0j}\mathcal{C}_{0z}\partial_{j}\partial^{i}\delta S^{0z}+\tilde{\chi}_{3}S_{(0)}^{0j}\mathcal{D}_{xy}\partial_{j}\partial^{i}\delta S^{xy}+\tilde{\chi}_{3}S_{(0)}^{0i}\chi_{T}\partial_{j}\partial_{j}\delta e \notag \\
  &\quad +\tilde{\chi}_{3}S_{(0)}^{0i}\mathcal{C}_{0z}\partial_{j}\partial_{j}\delta S^{0z}+\tilde{\chi}_{3}S_{(0)}^{0i}\mathcal{D}_{xy}\partial_{j}\partial_{j}\delta S^{xy}, \\
 0&=\partial_{0}\left(\delta S^{ij}\right)+S_{(0)}^{ij}\partial_{n}\left(\delta u^{n}\right)+\tilde{\chi}_{4}\beta_{(0)}\partial_{n}\partial^{n}\delta S^{ij}+\tilde{\chi}_{4}S_{(0)}^{ij}\chi_{T}\partial_{n}\partial^{n}\delta e \notag \\
  &\quad +\tilde{\chi}_{4}S_{(0)}^{ij}\mathcal{C}_{0z}\partial_{n}\partial^{n}\delta S^{0z}+\tilde{\chi}_{4}S_{(0)}^{ij}\mathcal{D}_{xy}\partial_{n}\partial^{n}\delta S^{xy},
\end{align}
where
\begin{gather}
 \vartheta^{i} \equiv \left(e_{(0)}+p_{(0)}\right)\delta u^{i}, \\
 \gamma_{\parallel} \equiv \frac{\frac{4}{3}\eta+\zeta}{e_{(0)}+p_{(0)}},\quad
 \gamma_{\perp} \equiv \frac{\eta}{e_{(0)}+p_{(0)}}.
\end{gather}

Next, the perturbations $\delta e$, $\delta u^\mu$, and $\delta S^{\mu\nu}$ in the plane-wave form are given as
\begin{align}\label{eq:18}
 \delta X=\delta\tilde{X}e^{i\omega t-i \vec{k}\cdot \vec{x}}.
\end{align}
In the absence of a spin density background, the system exhibits rotational symmetry and therefore any wave vector direction can be chosen for analysis. However, once a nonvanishing spin density background along the $z$ direction is assumed, $SO(3)$ symmetry is explicitly broken. As a consequence, the mode analyses for perturbations propagating parallel and perpendicular to the $z$ direction are no longer equivalent. The separate mode analyses for wave vectors along the $z$ direction $(k=(0,0,k_z))$ and along the $x$ direction $(k=(k_x,0,0))$  are investigated in this work. It should be emphasized that the stability and causality conditions, Eqs. \eqref{eq:8}–\eqref{eq:10}, are universal and do not rely on $\mathrm{SO}(3)$ symmetry. The stability criterion Eq. \eqref{eq:8} originates from the fundamental requirement of thermodynamic equilibrium, namely that all fluctuations must decay; hence this condition must hold regardless of spatial anisotropy. Similarly, the causality constraints Eqs. \eqref{eq:9}–\eqref{eq:10} stem from the principle of the constancy of the speed of light and must be satisfied for all cases.

  \subsubsection{Mode analysis along the $x$ direction}
We first consider the case in which the wave-vector is along the $x$ direction, i.e. $k=(k_x,0,0)$. With the plane-wave ansatz in Eq. \eqref{eq:18}, the spin hydrodynamic equations can be written as
\begin{align}
 \mathcal{M}_{1}\delta\tilde{X}_{1}=0,
\end{align}
where
\begin{gather}
 \mathcal{M}_{1}=\begin{pmatrix}M_{1} & 0\\
 0 & M_{2}
 \end{pmatrix}, \\
 \delta\tilde{X}_{1} \equiv \left(\delta\tilde{e},\delta\tilde{\vartheta}^{x},\delta\tilde{S}^{xy},\delta\tilde{S}^{0z},\delta\tilde{\vartheta}^{y},\delta\tilde{\vartheta}^{z},\delta\tilde{S}^{0x},\delta\tilde{S}^{0y},\delta\tilde{S}^{xz},\delta\tilde{S}^{yz}\right)^{\mathrm{T}},
\end{gather}
and
\begin{gather}
 M_{1}=\begin{pmatrix}
 i\omega & -ik & 0 & 0\\
 -ik\chi_{c} & i\omega+\gamma_{\parallel}k^{2} & -ik\mathcal{B}_{xy} & -ik\mathcal{A}_{0z}\\
 \tilde{\chi}_{4}\chi_{T}S_{(0)}^{xy}k^{2} & -ik\theta S_{(0)}^{xy} & i\omega+\tilde{\chi}_{4}\mathcal{D}_{T}k^{2} & \tilde{\chi}_{4}\mathcal{C}_{0z}S_{(0)}^{xy}k^{2}\\
 -\tilde{\chi}_{23}\chi_{T}S_{(0)}^{0z}k^{2} & -ik\theta S_{(0)}^{0z} & -\tilde{\chi}_{23}\mathcal{D}_{xy}S_{(0)}^{0z}k^{2} & i\omega-\tilde{\chi}_{23}\mathcal{C}_{T}k^{2}
 \end{pmatrix}, \\
 M_{2}=\begin{pmatrix}
 i\omega+\gamma_{\perp}k^{2} & 0 & 0 & 0 & 0 & 0\\
 0 & i\omega+\gamma_{\perp}k^{2} & 0 & 0 & 0 & 0\\
 0 & 0 & i\omega-\tilde{\chi}_{12}\beta_{(0)}k_{x}^{2} & 0 & 0 & 0\\
 0 & 0 & 0 & i\omega-\tilde{\chi}_{23}\beta_{(0)}k_{x}^{2} & 0 & 0\\
 0 & 0 & 0 & 0 & i\omega+\tilde{\chi}_{4}\beta_{(0)}k^{2} & 0\\
 0 & 0 & 0 & 0 & 0 & i\omega+\tilde{\chi}_{4}\beta_{(0)}k^{2}
 \end{pmatrix}. \label{eq:30}
\end{gather}

The dispersion relations are determined by the existence of nontrivial solutions of the characteristic matrix $\mathcal{M}_{1}$, i.e.,~$\mathrm{det}\mathcal{M}_{1}=0$.

In this case, the characteristic matrix $\mathcal{M}_{1}$ is diagonal, and the condition can be simplified to
\begin{align}
 \mathrm{det}M_{1}\cdot\mathrm{det}M_{2}=0.
\end{align}
Since $M_{2}$ is diagonal, the corresponding dispersion relations $\mathrm{det}M_{2}=0$ can be obtained straightforwardly as
\begin{align}
 \omega&=i\gamma_{\perp}k^{2}, \label{eq:19} \\
 \omega&=-i\beta_{(0)}\tilde{\chi}_{12}k^{2},  \label{eq:20} \\
 \omega&=-i\beta_{(0)}\tilde{\chi}_{23}k^{2},  \label{eq:21} \\
 \omega&=i\beta_{(0)}\tilde{\chi}_{4}k^{2}.  \label{eq:22}
\end{align}

Equation \eqref{eq:19} represents the shearlike modes, whose damping is governed solely by $\gamma_{\perp}$ and remains unaffected by the spin background at leading order. In contrast, Eqs.  \eqref{eq:20}-\eqref{eq:22} are spin dissipative modes, corresponding to the relaxation of spin currents such as $\Sigma$, $\Sigma_{(s)}^{\left\langle \mu\alpha\right\rangle }$, $\Sigma_{(a)}^{[\mu\alpha]}$ and $\Sigma^{\left\langle \mu\right\rangle \left\langle \alpha\right\rangle \left\langle \beta\right\rangle }$, reflecting dissipative processes intrinsic to the spin sector. Notably, these modes persist even in the zero-spin background. This is because they are independent of the state parameters of the nonvanishing spin background ($\mathcal{A}$, $\mathcal{B}$, etc.).

For $M_{1}$, a direct analytical treatment is rather involved; therefore, we focus on the dispersion relations in various limiting cases. In the small wave-vector limit $k\rightarrow0$,  one obtains
\begin{align}\label{eq:23}
 \omega=\pm c_{v}k+i\left(\frac{1}{2}\gamma_{\parallel}+\frac{\left[\chi_{T}+\theta\left(\beta_{(0)}+\mathcal{C}+\mathcal{D}\right)\right]\left(\tilde{\chi}_{4}\mathcal{B}-\tilde{\chi}_{23}\mathcal{A}\right)}{2c^{2}_{v}}\right)k^{2},
\end{align}
where the introduced parameter $c_{v}$ is
\begin{gather}\label{eq:99}
 c_v=\sqrt{\chi_{c}+\frac{\mathcal{A}}{e_{(0)}+p_{(0)}}+\frac{\mathcal{B}}{e_{(0)}+p_{(0)}}}.
\end{gather}
The dispersion relation in Eq. \eqref{eq:23} corresponds to a complex coupled soundlike mode. The mode represents a propagating excitation with dissipative attenuation, analogous to the ordinary sound mode in an unpolarized fluid. Compared with the zero-spin background case ($\omega=\pm c_{s}k+i\frac{1}{2}\gamma_{\parallel}k^{2}$), however, the finite spin density background significantly modifies both the propagation and damping properties of the mode. The propagation velocity is no longer given by the conventional sound speed $c_s$, but is replaced by the effective velocity $c_v$, which depends explicitly on the spin background parameters $\mathcal{A}$ and $\mathcal{B}$. As a result, the spin background directly affects the causality condition through the value of $c_v$. In addition, the attenuation coefficient is no longer determined solely by the viscous parameter $\gamma_{\parallel}$. Instead, it receives additional contributions from the spin dissipative transport coefficients ($\tilde{\chi}_{4}$, $\tilde{\chi}_{23}$ or $\tilde{\chi}_{12}$) and the background spin density parameters ($\mathcal{A}$, $\mathcal{B}$, $\mathcal{C}$ and $\mathcal{D}$), indicating that the coupling between longitudinal hydrodynamic fluctuations and spin degrees of freedom can either enhance or suppress the damping of the sound mode. Therefore, while the mode retains its soundlike character, its effective propagation and dissipation are substantially altered by the finite spin background.

For the stability analysis, we apply the constraint Eq. \eqref{eq:8} to the dispersion relations of Eqs. \eqref{eq:19}-\eqref{eq:22} and \eqref{eq:23} in the small wave-vector limit, thereby obtaining the stability conditions as
\begin{align}\label{eq:96}
 \gamma_{\perp}>0,\quad \chi_{b}<0,\quad \chi_{s}>0,
\end{align}
\begin{align}\label{eq:97}
 \frac{1}{2}\gamma_{\parallel}+\frac{\left[\chi_{T}+\theta\left(\beta_{(0)}+\mathcal{C}+\mathcal{D}\right)\right]\left(\tilde{\chi}_{4}\mathcal{B}-\tilde{\chi}_{23}\mathcal{A}\right)}{2c^{2}_{v}}>0.
\end{align}
Applying the causality constraints Eqs. \eqref{eq:9} and \eqref{eq:10} to the dispersion relation yields the causality conditions
\begin{align}\label{eq:98}
 0\leq c_v\leq1.
\end{align}

It can be seen from Eq. \eqref{eq:96} that the constraints imposed by the stability of the system on the state parameters $\chi_{b}$ and $\chi_{s}$ are identical to those obtained in the zero-spin background case, even when a background spin density is present. The primary impact of the nontrivial spin background is reflected in the propagation mode Eq. \eqref{eq:23}.

In the large wave-vector limit $k\rightarrow\infty$, one obtains
\begin{gather}
 \omega=i\frac{\left(\beta_{(0)}+\mathcal{C}+\mathcal{D}\right)\chi_{c}-\left(\mathcal{A}+\mathcal{B}\right)\chi_{T}}{\gamma_{\parallel}\left(\beta_{(0)}+\mathcal{C}+\mathcal{D}\right)}, \label{eq:25} \\
 \omega=\frac{1}{2}i\left(\overline{a}_{x}\pm\sqrt{\overline{a}_{x}^{2}+\overline{b}_{x}}\right)k^{2}, \label{eq:24}
\end{gather}
where the introduced parameters $\overline{a}_{x}$ and $\overline{b}_{x}$ are given as
\begin{align}
 \overline{a}_{x}&=\left(\beta_{(0)}+\mathcal{D}\right)\tilde{\chi}_{4}-\left(\beta_{(0)}+\mathcal{C}\right)\tilde{\chi}_{23},  \\
 \overline{b}_{x}&=4\tilde{\chi}_{4}\tilde{\chi}_{23}\beta_{(0)}\left(\beta_{(0)}+\mathcal{C}+\mathcal{D}\right).
\end{align}

Modes \eqref{eq:25} and \eqref{eq:24} correspond to complex coupled spin dissipative modes arising from the mixing of different spin-current fluctuations in the presence of a finite spin background. Their damping rates are determined jointly by the spin transport coefficients and the spin background corrections, demonstrating that the relaxation dynamics of the spin sector is strongly influenced by both dissipative transport and background polarization effects. Equation \eqref{eq:25} represents a nonpropagating relaxation mode with a finite damping rate, whereas Eq. \eqref{eq:24} consists of two purely dissipative branches whose frequencies scale as $k^{2}$ in the ultraviolet limit.

Although Eq. \eqref{eq:24} does not describe propagating excitations, its ultraviolet behavior has important implications for causality. Since $\mathrm{Re}(\omega)=0$, the conditions $\underset{k\rightarrow\infty}{\lim}|\mathrm{Re}(\omega/k)|\le1$ and $\underset{k\rightarrow\infty}{\lim}|\mathrm{Re}(\partial\omega/\partial k)|\le1$ are trivially satisfied. However, Eq. \eqref{eq:24} scales as $\omega\propto i k^{2}$, implying $\left|\frac{\omega}{k}\right|\propto k$, which diverges as $k\rightarrow\infty$. Therefore, the boundedness requirement of the causality criterion is violated, and the mode is acausal. This behavior is analogous to the diffusive modes of conventional first-order hydrodynamics and the first-order spin hydrodynamic theory in the Landau frame. Consequently, the presence of a finite spin background does not remove the intrinsic acausality associated with first-order dissipative theories, although it modifies the corresponding damping rates through the spin-dependent coefficients $\mathcal{C}$ and $\mathcal{D}$.

\subsubsection{Mode analysis along the $z$ direction}
Similar to the case along the $x$ direction, the same conclusion can also be obtained when the wave vector is along the $z$ direction
\begin{align}
 \mathcal{M}_{2}\delta\tilde{X}_{2}=0,
\end{align}
where
\begin{gather}
 \mathcal{M}_{2}=\begin{pmatrix}M_{3} & 0\\
 0 & M_{4}
\end{pmatrix}, \\
 \delta\tilde{X}_{2} \equiv \left(\delta\tilde{e},\delta\tilde{\vartheta}^{z},\delta\tilde{S}^{xy},\delta\tilde{S}^{0z},\delta\tilde{\vartheta}^{y},\delta\tilde{\vartheta}^{x},\delta\tilde{S}^{0x},\delta\tilde{S}^{0y},\delta\tilde{S}^{xz},\delta\tilde{S}^{yz}\right)^{\mathrm{T}},
\end{gather}
where
\begin{gather}
 M_{3}=\begin{pmatrix}
 i\omega & -ik & 0 & 0\\
 -ik\chi_{c} & i\omega+\gamma_{\parallel}k^{2} & -ik\mathcal{B}_{xy} & -ik\mathcal{A}_{0z}\\
 \tilde{\chi}_{4}\chi_{T}S_{(0)}^{xy}k^{2} & -ik\theta S_{(0)}^{xy} & i\omega+\mathcal{D}_{T}\tilde{\chi}_{4}k^{2} & \tilde{\chi}_{4}\mathcal{C}_{0z}S_{(0)}^{xy}k^{2}\\
 -\tilde{\chi}_{12}\chi_{T}S_{(0)}^{0z}k^{2} & -ik\theta S_{(0)}^{0z} & -\tilde{\chi}_{12}\mathcal{D}_{xy}S_{(0)}^{0z}k^{2} & i\omega-\tilde{\chi}_{12}\mathcal{C}_{T}k^{2}
 \end{pmatrix}, \\
 M_{4}=\begin{pmatrix}
 i\omega+\gamma_{\perp}k^{2} & 0 & 0 & 0 & 0 & 0\\
 0 & i\omega+\gamma_{\perp}k^{2} & 0 & 0 & 0 & 0\\
 0 & 0 & i\omega-\tilde{\chi}_{23}\beta_{(0)}k^{2} & 0 & 0 & 0\\
 0 & 0 & 0 & i\omega-\tilde{\chi}_{23}\beta_{(0)}k^{2} & 0 & 0\\
 0 & 0 & 0 & 0 & i\omega+\tilde{\chi}_{4}\beta_{(0)}k^{2} & 0\\
 0 & 0 & 0 & 0 & 0 & i\omega+\tilde{\chi}_{4}\beta_{(0)}k^{2}
 \end{pmatrix}. \label{eq:31}
\end{gather}

From $\mathrm{det}M_{4}=0$, one can obtain the following relationship as
\begin{gather}
 \omega=i\gamma_{\perp}k^{2}, \label{eq:27} \\
 \omega=-i\tilde{\chi}_{23}\beta_{(0)}k^{2}, \label{eq:28}\\
 \omega=i\tilde{\chi}_{4}\beta_{(0)}k^{2}. \label{eq:29}
\end{gather}

In the small wave vector limit of $k\rightarrow0$ , from $\mathrm{det}M_{3}=0$, one can obtain
\begin{align}\label{eq:100}
\omega=\pm c_{v}k+i\left(\frac{1}{2}\gamma_{\parallel}+\frac{\left[\chi_{T}+\theta\left(\beta_{(0)}+\mathcal{C}+\mathcal{D}\right)\right]\left(\tilde{\chi}_{4}\mathcal{B}-\tilde{\chi}_{12}\mathcal{A}\right)}{2c^{2}_{v}}\right)k^{2}.
\end{align}

In the large wave-vector limit $k\rightarrow\infty$, one obtains
\begin{gather}
 \omega=i\frac{\left(\beta_{(0)}+\mathcal{C}+\mathcal{D}\right)\chi_{c}-\left(\mathcal{A}+\mathcal{B}\right)\chi_{T}}{\gamma_{\parallel}\left(\beta_{(0)}+\mathcal{C}+\mathcal{D}\right)},\label{eq:106}\\
 \omega=\frac{1}{2}i\left(\overline{a}_{z}\pm\sqrt{\overline{a}_{z}^{2}+\overline{b}_{z}}\right)k^{2} \label{eq:101},
\end{gather}
where
\begin{align}
 \overline{a}_{z}&=\left(\beta_{(0)}+\mathcal{D}\right)\tilde{\chi}_{4}-\left(\beta_{(0)}+\mathcal{C}\right)\tilde{\chi}_{12},  \\
  \overline{b}_{z}&=4\tilde{\chi}_{4}\tilde{\chi}_{12}\beta_{(0)}\left(\beta_{(0)}+\mathcal{C}+\mathcal{D}\right).
\end{align}
The dispersion relations along the $z$ direction are given by Eqs. \eqref{eq:27}-\eqref{eq:29} and \eqref{eq:100}-\eqref{eq:101}. 

The dispersion relations in the $z$ direction exhibit the same qualitative structure as those obtained for perturbations propagating along the $x$ direction. Specifically, Eq. \eqref{eq:27} corresponds to a shear-like mode, while Eqs. \eqref{eq:28} and \eqref{eq:29} describe nonpropagating spin dissipative modes associated with the relaxation of spin-current fluctuations. Equation \eqref{eq:100} represents a complex coupled soundlike propagating mode, whereas Eqs. \eqref{eq:106} and \eqref{eq:101}, together with the accompanying non-propagating branch, correspond to complex coupled spin dissipative modes in the large wave-vector limit. Therefore, the classification of the modes remains unchanged when the propagation direction is rotated from the transverse ($x$) direction to the longitudinal ($z$) direction. Regarding the stability and causality conditions, they share the same structure as the their counterparts in the $x$ direction which are presented as

\begin{align}
 \gamma_{\perp}>0,\quad \chi_{b}<0,\quad \chi_{s}>0,
\end{align}
\begin{align}
 \frac{1}{2}\gamma_{\parallel}+\frac{\left[\chi_{T}+\theta\left(\beta_{(0)}+\mathcal{C}+\mathcal{D}\right)\right]\left(\tilde{\chi}_{4}\mathcal{B}-\tilde{\chi}_{12}\mathcal{A}\right)}{2c^{2}_{v}}>0,
\end{align}
\begin{align}
 0\leq c_v\leq1.
\end{align}
Similarly, Eq. \eqref{eq:101} also leads to acausality.

Although the mode classification is unchanged, quantitative differences arise between the transverse ($x$ direction) and longitudinal ($z$ direction) propagation directions. The complex coupled soundlike modes in Eqs. \eqref{eq:23} and \eqref{eq:100} possess identical real parts, implying that the propagation velocity of long-wavelength disturbances is independent of the direction relative to the spin background. However, their imaginary parts (damping terms) differ, implying that the decay rates of perturbations along the $x$ and $z$ directions are distinct. This difference arises from different combinations of spin dissipative transport coefficients: the damping is governed by $\chi_{23}$ in the $x$ direction and by $\chi_{12}$ in the $z$ direction. The same relation holds between modes of Eqs. \eqref{eq:24} and \eqref{eq:101}. Furthermore, compared with the $x$ direction, the purely dissipative modes in the $z$ direction lack the mode of Eq. \eqref{eq:20}, indicating that this mode exists only in specific directions and exhibits directional dependence. All these features reflect the breaking of the $SO(3)$ symmetry of the system.

Combining the dispersion relations in both directions, we conclude that within our framework, the first-order spin fluid with a nonvanishing spin background is acausal, as in previous first-order fluid theories. To obtain results consistent with fundamental physical constraints, we should extend the first-order fluid theory to the second-order theory.

\section{Linear Modal Analysis of Second-Order Minimal Causal Spin Hydrodynamics}\label{chap:3}
 \subsection{Second-order relativistic spin hydrodynamics}
The stability and causality of first-order spin hydrodynamics in the presence of a nonvanishing spin density background were investigated in the previous section. The results show that the first-order theory exhibits acausal behavior and thus cannot describe realistic physical systems. This is consistent with the known fact that first-order theories in relativistic hydrodynamics are generically acausal \cite{Hiscock:1985zz,Hiscock:1987zz,Daher:2022wzf,Xie:2023gbo}. However, recent studies suggest that stable and causal first-order theories might exist within generalized frameworks, as demonstrated by the Bemfica-Disconzi-Noronha-Kovtun (BDNK) framework \cite{Armas:2022wvb,Kovtun:2019hdm,Bemfica:2022dnk}. For further study on the connection between BDNK theory and second-order fluid theories, refer to Ref. \cite{Noronha:2021syv}.

To construct a stable and causal hydrodynamic theory, one may consider second-order formulations \cite{Israel:1979wp,10.1098/rspa.1979.0005,Baier:2007ix,Denicol:2012cn}. In this section, the M$\ddot{\textrm{u}}$ller-Israel-Stewart (MIS) theory is adopted, whose core idea is to incorporate second-order corrections into a first-order framework. Among all second-order terms, those proportional to relaxation time play a crucial role in restoring causality. When these terms are included, the hydrodynamic equations transition from parabolic to hyperbolic type, thereby allowing causal signal propagation. To simplify the analysis, we therefore only consider second-order corrections involving relaxation time, leading to what is called minimal causal spin hydrodynamics \cite{Xie:2023gbo,Koide:2006ef}. For studies incorporating complete noncrossed second-order terms, see Refs. \cite{Daher:2024bah,Biswas:2023qsw}. The constitutive relations now take the form of
\begin{gather}
 \tau_{\Pi}\dfrac{d}{d\tau}\Pi+\Pi=-\zeta\partial_{\mu}u^{\mu}, \\
 \tau_{\pi}\Delta^{\alpha\langle\mu}\Delta^{\nu\rangle\beta}\dfrac{d}{d\tau}\pi_{\alpha\beta}+\pi^{\mu\nu}=2\eta\partial^{\langle\mu}u^{\nu\rangle}, \\
 \tau_{\Sigma}\dfrac{d}{d\tau}\Sigma+\Sigma=-\chi_{1}\left[u^{\beta}\Delta^{\mu\alpha}\partial_{\mu}\left(\beta\omega_{\alpha\beta}\right)\right], \\
 \tau_{\Sigma_{s}}\Delta^{\gamma\langle\mu}\Delta^{\alpha\rangle\beta}\dfrac{d}{d\tau}\Sigma_{(s)\gamma\beta}+\Sigma_{(s)}^{\left\langle \mu\alpha\right\rangle }=-\chi_{2}u_{\beta}\left(\Delta_{\ \gamma}^{\mu}\Delta_{\ \sigma}^{\alpha}+\Delta_{\ \sigma}^{\mu}\Delta_{\ \gamma}^{\alpha}-\frac{2}{3}\Delta^{\mu\alpha}\Delta_{\gamma\sigma}\right)\partial^{\gamma}\left(\beta\omega^{\sigma\beta}\right), \\
 \tau_{\Sigma_{a}}\Delta^{\mu\gamma}\Delta^{\alpha\beta}\dfrac{d}{d\tau}\Sigma_{(a)\gamma\beta}+\Sigma_{(a)}^{[\mu\alpha]}=-\chi_{3}u^{\beta}\Delta^{\mu\zeta}\Delta^{\alpha\lambda}\partial_{[\zeta}\left(\beta\omega_{\lambda]\beta}\right), \\
 \tau_{\Sigma_{4}}\Delta^{\mu\rho}\Delta^{\alpha[\eta}\Delta^{\beta,\kappa]}\dfrac{d}{d\tau}\Sigma_{\left\langle \rho\right\rangle \left\langle \eta\right\rangle \left\langle \kappa\right\rangle }+\Sigma^{\left\langle \mu\right\rangle \left\langle \alpha\right\rangle \left\langle \beta\right\rangle }=\chi_{4}\Delta^{\mu\rho}\Delta^{\alpha\eta}\Delta^{\beta\kappa}\partial_{\rho}\left(\beta\omega_{\eta\kappa}\right),
\end{gather}
where $\tau_{\Pi}$, $\tau_{\pi}$, $\tau_{\Sigma}$, $\tau_{\Sigma_{s}}$, $\tau_{\Sigma_{a}}$ and $\tau_{\Sigma_{4}}$ denote the relaxation times of the corresponding dissipative currents, and all these quantities are positive definite.

In the second-order theory, all dissipative quantities will be treated as independent variables like the main hydrodynamic variables. So the hydrodynamic perturbation equations are given as
\begin{align}
 0&=\partial_{0}\left(\delta e\right)+\partial_{i}\left(\delta\vartheta^{i}\right), \\
 0&=-\chi_{c}\partial^{j}\delta e+\partial_{0}\delta\vartheta^{j}-\mathcal{A}_{0z}\partial^{j}\delta S^{0z}-\mathcal{B}_{xy}\partial^{j}\delta S^{xy}-\partial^{j}\delta\Pi+\partial_{i}\left(\delta\pi^{ij}\right), \\
 0&=\partial_{0}\left(\delta S^{0i}\right)+\theta S_{(0)}^{0i}\partial_{j}\left(\delta\vartheta^{j}\right)-\partial^{i}\delta\Sigma-\partial_{j}\delta\Sigma_{(s)}^{\left\langle ij\right\rangle }+\partial_{j}\delta\Sigma_{(a)}^{[ij]}, \\
 0&=\partial_{0}\left(\delta S^{ij}\right)+\theta S_{(0)}^{ij}\partial_{n}\left(\delta\vartheta^{n}\right)+\partial_{n}\left(\delta\Sigma^{\left\langle n\right\rangle \left\langle i\right\rangle \left\langle j\right\rangle }\right), \\
 \delta\Pi
  &=-\left(\gamma_{\parallel}-\frac{4}{3}\gamma_{\perp}\right)\partial_{i}\delta\vartheta^{i}-\tau_{\Pi}\partial_{0}\delta\Pi, \\
 \delta\pi^{ij}
  &=\gamma_{\perp}\left[\partial^{i}\delta\vartheta^{j}+\partial^{j}\delta\vartheta^{i}-\frac{2}{3}g^{ij}\partial_{\alpha}\delta\vartheta^{\alpha}\right]-\tau_{\pi}D\delta\pi^{ij}, \\
 \delta\Sigma
  &=\tilde{\chi}_{1}\chi_{T}S_{(0)0j}\partial^{j}\delta e+\tilde{\chi}_{1}\mathcal{C}_{0z}S_{(0)0j}\partial^{j}\delta S^{0z}+\tilde{\chi}_{1}\mathcal{D}_{xy}S_{(0)0j}\partial^{j}\delta S^{xy} \notag \\
   &\quad +\tilde{\chi}_{1}\beta_{(0)}\partial_{j}\delta S^{0j}-\tau_{\Sigma}\partial_{0}\delta\Sigma, \\
 \delta\Sigma_{(s)}^{\left\langle ij\right\rangle }
  &=-\tau_{\Sigma_{s}}\partial_{0}\delta\Sigma_{(s)}^{ij}+\tilde{\chi}_{2}\beta_{(0)}\partial^{i}\delta S^{0j}+\tilde{\chi}_{2}\beta_{(0)}\partial^{j}\delta S^{0i}-\frac{2}{3}\tilde{\chi}_{2}\beta_{(0)}g_{(0)}^{ij}\partial_{m}\delta S^{0m} \notag \\
   &\quad +\tilde{\chi}_{2}S_{(0)}^{0j}\chi_{T}\partial^{i}\delta e+\tilde{\chi}_{2}S_{(0)}^{0j}\mathcal{C}_{0z}\partial^{i}\delta S^{0z}+\tilde{\chi}_{2}S_{(0)}^{0j}\mathcal{D}_{xy}\partial^{i}\delta S^{xy} \notag \\
   &\quad +\tilde{\chi}_{2}S_{(0)}^{0i}\chi_{T}\partial^{j}\delta e+\tilde{\chi}_{2}S_{(0)}^{0i}\mathcal{C}_{0z}\partial^{j}\delta S^{0z}+\tilde{\chi}_{2}S_{(0)}^{0i}\mathcal{D}_{xy}\partial^{j}\delta S^{xy}  \notag \\
   &\quad -\frac{2}{3}\tilde{\chi}_{2}g_{(0)}^{ij}S_{(0)}^{0m}\chi_{T}\partial_{m}\delta e-\frac{2}{3}\tilde{\chi}_{2}g_{(0)}^{ij}S_{(0)}^{0m}\mathcal{C}_{0z}\partial_{m}\delta S^{0z}-\frac{2}{3}\tilde{\chi}_{2}g_{(0)}^{ij}S_{(0)}^{0m}\mathcal{D}_{xy}\partial_{m}\delta S^{xy}, \\
 \delta\Sigma_{(a)}^{[ij]}
  &=-\tau_{\Sigma_{a}}\partial_{0}\delta\Sigma_{(a)}^{ij}+\tilde{\chi}_{3}\beta_{(0)}\partial^{i}\delta S^{0j}-\tilde{\chi}_{3}\beta_{(0)}\partial^{j}\delta S^{0i} \notag \\
   &\quad +\tilde{\chi}_{3}S_{(0)}^{0j}\chi_{T}\partial^{i}\delta e+\tilde{\chi}_{3}S_{(0)}^{0j}\mathcal{C}_{0z}\partial^{i}\delta S^{0z}+\tilde{\chi}_{3}S_{(0)}^{0j}\mathcal{D}_{xy}\partial^{i}\delta S^{xy} \notag \\
   &\quad -\tilde{\chi}_{3}S_{(0)}^{0i}\chi_{T}\partial^{j}\delta e-\tilde{\chi}_{3}S_{(0)}^{0i}\mathcal{C}_{0z}\partial^{j}\delta S^{0z}-\tilde{\chi}_{3}S_{(0)}^{0i}\mathcal{D}_{xy}\partial^{j}\delta S^{xy}, \\
 \delta\Sigma^{\left\langle n\right\rangle \left\langle i\right\rangle \left\langle j\right\rangle }
 &=-\tau_{\Sigma_{4}}\partial_{0}\delta\Sigma^{\left\langle n\right\rangle \left\langle i\right\rangle \left\langle j\right\rangle }+\tilde{\chi}_{4}\beta_{(0)}\partial^{n}\delta S^{ij}+\tilde{\chi}_{4}S_{(0)}^{ij}\chi_{T}\partial^{n}\delta e \notag \\
  &\quad +\tilde{\chi}_{4}S_{(0)}^{ij}\mathcal{C}_{0z}\partial^{n}\delta S^{0z}+\tilde{\chi}_{4}S_{(0)}^{ij}\mathcal{D}_{xy}\partial^{n}\delta S^{xy}.
\end{align}

 \subsection{Linear mode analysis at second order}
  \subsubsection{Mode analysis along the $x$ direction}

By first analyzing the situation in the $x$ direction, one can obtain the hydrodynamic equations as
\begin{align}
\mathcal{M}_{3}\delta\tilde{X}_{3}=0,
\end{align}
where
\begin{align}
 \mathcal{M}_{3}=
 \begin{pmatrix}
 M_{5} & 0 & 0 & 0 & 0 & 0\\
 0 & M_{6} & 0 & 0 & 0 & 0\\
 0 & 0 & M_{7} & 0 & 0 & 0\\
 0 & 0 & 0 & M_{8} & 0 & 0\\
 0 & 0 & 0 & 0 & M_{9} & 0\\
 0 & 0 & 0 & 0 & 0 & M_{10}
 \end{pmatrix},
\end{align}
\begin{align}
\delta\tilde{X}_{3}
=(&\delta\tilde{e},\delta\tilde{\vartheta}^{x},\delta\tilde{S}^{0x},\delta\tilde{\Pi},\delta\tilde{\Sigma},\delta\tilde{\pi}^{xx},\delta\tilde{\Sigma}_{(s)}^{xx},\delta\tilde{\vartheta}^{y},\delta\tilde{S}^{0y},\delta\tilde{S}^{xy}, \notag \\
& \delta\tilde{\pi}^{xy},\delta\tilde{\Sigma}_{(s)}^{xy},\delta\tilde{\Sigma}_{(a)}^{xy},\delta\tilde{\Sigma}^{xxy},\delta\tilde{\vartheta}^{z},\delta\tilde{S}^{0z},\delta\tilde{S}^{xz},\delta\tilde{\pi}^{xz}, \notag \\
& \delta\tilde{\Sigma}_{(s)}^{xz},\delta\tilde{\Sigma}_{(a)}^{xz},\delta\tilde{\Sigma}^{xxz},\delta\tilde{S}^{yz},\delta\tilde{\pi}^{yz},\delta\tilde{\Sigma}_{(s)}^{yz},\delta\tilde{\Sigma}_{(a)}^{yz},\delta\tilde{\Sigma}^{xyz}, \notag \\
& \delta\tilde{\pi}^{yy},\delta\tilde{\Sigma}_{(s)}^{yy},\delta\tilde{\Sigma}^{yxy},\delta\tilde{\Sigma}^{zxy},\delta\tilde{\Sigma}^{yxz},\delta\tilde{\Sigma}^{zxz},\delta\tilde{\Sigma}^{yyz},\delta\tilde{\Sigma}^{zyz})^T.
\end{align}

The explicit expressions for each block matrix in $\mathcal{M}_{3}$ are
\begin{gather}
M_{5}=
\begin{pmatrix}i\omega & -ik & 0 & 0 & 0 & 0 & 0 & 0 & 0 & 0\\
-ik\chi_{c} & i\omega & 0 & 0 & 0 & -ik\mathcal{A}_{0z} & -ik\mathcal{B}_{xy} & -ik & -ik & 0\\
ik\tilde{\chi}_{2}S_{(0)}^{0z}\chi_{T} & 0 & \Gamma_{4} & 0 & 0 & ik\tilde{\chi}_{2}\mathcal{C}_{T} & ik\tilde{\chi}_{2}\mathcal{D}_{xy}S_{(0)}^{0z} & 0 & 0 & 0\\
ik\tilde{\chi}_{3}S_{(0)}^{0z}\chi_{T} & 0 & 0 & \Gamma_{5} & 0 & ik\tilde{\chi}_{3}\mathcal{C}_{T} & ik\tilde{\chi}_{3}\mathcal{D}_{xy}S_{(0)}^{0z} & 0 & 0 & 0\\
ik\tilde{\chi}_{4}S_{(0)}^{xy}\chi_{T} & 0 & 0 & 0 & \Gamma_{6} & ik\tilde{\chi}_{4}\mathcal{C}_{0z}S_{(0)}^{xy} & ik\tilde{\chi}_{4}\mathcal{D}_{T} & 0 & 0 & 0\\
0 & -ik\theta S_{(0)}^{0z} & ik & ik & 0 & i\omega & 0 & 0 & 0 & 0\\
0 & -ik\theta S_{(0)}^{xy} & 0 & 0 & -ik & 0 & i\omega & 0 & 0 & 0\\
0 & -\left(\gamma_{\parallel}-\frac{4}{3}\gamma_{\perp}\right)ik & 0 & 0 & 0 & 0 & 0 & -\Gamma_{2} & 0 & 0\\
0 & -\frac{4}{3}\gamma_{\perp}ik & 0 & 0 & 0 & 0 & 0 & 0 & -\Gamma_{1} & 0\\
0 & \frac{2}{3}\gamma_{\perp}ik & 0 & 0 & 0 & 0 & 0 & 0 & 0 & -\Gamma_{1}
\end{pmatrix},
\end{gather}

\begin{align}
 M_{6}=
 \begin{pmatrix}i\omega & 0 & -ik & 0\\
 0 & i\omega & 0 & -ik\\
 \gamma_{\perp}ik & 0 & \Gamma_{1} & 0\\
 0 & \gamma_{\perp}ik & 0 & \Gamma_{1}
 \end{pmatrix},
\end{align}

\begin{align}
 M_{7}=
 \begin{pmatrix}i\omega & -ik & ik & 0\\
 -ik\tilde{\chi}_{1}\beta_{(0)} & \Gamma_{3} & 0 & 0\\
 \frac{4}{3}ik\tilde{\chi}_{2}\beta_{(0)} & 0 & \Gamma_{4} & 0\\
 -\frac{2}{3}ik\tilde{\chi}_{2}\beta_{(0)} & 0 & 0 & \Gamma_{4}
 \end{pmatrix},
\end{align}

\begin{align}
 M_{8}=
 \begin{pmatrix}i\omega & ik & ik\\
 ik\tilde{\chi}_{2}\beta_{(0)} & \Gamma_{3} & 0\\
 ik\tilde{\chi}_{3}\beta_{(0)} & 0 & \Gamma_{5}
 \end{pmatrix},
\end{align}

\begin{align}
 M_{9}=
 \begin{pmatrix}i\omega & 0 & -ik & 0\\
 0 & i\omega & 0 & -ik\\
 ik\tilde{\chi}_{4}\beta_{(0)} & 0 & \Gamma_{6} & 0\\
 0 & ik\tilde{\chi}_{4}\beta_{(0)} & 0 & \Gamma_{6}
 \end{pmatrix},
\end{align}

\begin{align}
 M_{10}=
 \begin{pmatrix}\Gamma_{1} & 0 & 0 & 0 & 0 & 0 & 0 & 0 & 0\\
 0 & \Gamma_{4} & 0 & 0 & 0 & 0 & 0 & 0 & 0\\
 0 & 0 & \Gamma_{5} & 0 & 0 & 0 & 0 & 0 & 0\\
 0 & 0 & 0 & \Gamma_{6} & 0 & 0 & 0 & 0 & 0\\
 0 & 0 & 0 & 0 & \Gamma_{6} & 0 & 0 & 0 & 0\\
 0 & 0 & 0 & 0 & 0 & \Gamma_{6} & 0 & 0 & 0\\
 0 & 0 & 0 & 0 & 0 & 0 & \Gamma_{6} & 0 & 0\\
 0 & 0 & 0 & 0 & 0 & 0 & 0 & \Gamma_{6} & 0\\
 0 & 0 & 0 & 0 & 0 & 0 & 0 & 0 & \Gamma_{6}
 \end{pmatrix}.
\end{align}
where
\begin{align}
  \begin{array}{ccc}
  \Gamma_{1}=-\tau_{\pi}i\omega-1, & \Gamma_{2}=-\tau_{\Pi}i\omega-1, & \Gamma_{3}=-i\omega\tau_{\Sigma}-1,
  \end{array} \notag \\
  \begin{array}{ccc}
  \Gamma_{4}=-i\omega\tau_{\Sigma_{s}}-1, & \Gamma_{5}=-i\omega\tau_{\Sigma_{a}}-1, & \Gamma_{6}=-i\omega\tau_{\Sigma_{4}}-1.
  \end{array}
\end{align}

From $\mathrm{det}M_{10}=0$ , one can obtain
\begin{align}
 \omega&=\frac{i}{\tau_{\pi}}, \\
 \omega&=\frac{i}{\tau_{\Sigma_{s}}}, \\
 \omega&=\frac{i}{\tau_{\Sigma_{a}}}, \\
 \omega&=\frac{i}{\tau_{\Sigma_{4}}}.
\end{align}

One can obtain the results from $\mathrm{det}M_{j}=0~(j\neq10)$ in the $k\rightarrow0$ limit as
\begin{gather}
 \omega=\frac{i}{\tau_{\Sigma}}, \label{eq:32} \\
 \omega=\frac{i}{\tau_{\pi}}, \label{eq:33} \\
 \omega=\frac{i}{\tau_{\Pi}}, \label{eq:34} \\
 \omega=\frac{i}{\tau_{\Sigma_{s}}}, \label{eq:35} \\
 \omega=\frac{i}{\tau_{\Sigma_{a}}}, \label{eq:36} \\
 \omega=\frac{i}{\tau_{\Sigma_{4}}}, \label{eq:37} \\
 \omega=i\gamma_{\perp}k^{2}, \label{eq:38} \\
 \omega=-i\beta_{(0)}\tilde{\chi}_{12}k^{2}, \label{eq:39} \\
 \omega=-i\beta_{(0)}\tilde{\chi}_{23}k^{2}, \label{eq:40} \\
 \omega=i\beta_{(0)}\tilde{\chi}_{4}k^{2}, \label{eq:41} \\
 \omega=\pm c_{v}k+i\left(\frac{1}{2}\gamma_{\parallel}+\frac{\left[\chi_{T}+\theta\left(\beta_{(0)}+\mathcal{C}+\mathcal{D}\right)\right]\left(\tilde{\chi}_{4}\mathcal{B}-\tilde{\chi}_{23}\mathcal{A}\right)}{2c_{v}^{2}}\right)k^{2}, \label{eq:42}
\end{gather}
where $\mathrm{det}M_{5}=0$ gives Eqs. \eqref{eq:33}-\eqref{eq:36} and \eqref{eq:42}; $\mathrm{det}M_{6}=0$ gives Eqs.  \eqref{eq:34} and \eqref{eq:38}; $\mathrm{det}M_{7}=0$ gives Eqs. \eqref{eq:32}, \eqref{eq:35} and \eqref{eq:39}; $\mathrm{det}M_{8}=0$  gives Eqs. \eqref{eq:35}, \eqref{eq:36} and \eqref{eq:40}; $\mathrm{det}M_{9}=0$ gives Eqs. \eqref{eq:37} and \eqref{eq:41}. 
Compared with the first-order theory, the most notable new feature in the long-wavelength limit is the appearance of the relaxation modes in Eqs. \eqref{eq:32}--\eqref{eq:37}, whose frequencies are determined solely by the corresponding relaxation times. These modes originate from promoting dissipative currents to independent dynamical variables within the second-order framework and therefore have no counterparts in the first-order theory.

Apart from these additional relaxation modes, the remaining hydrodynamic spectrum, Eqs. \eqref{eq:38}--\eqref{eq:42}, retains the same structure as in the first-order spin fluid. Consequently, the classification and physical interpretation of the shearlike, spin dissipative, and complex coupled soundlike modes remain unchanged. This demonstrates that the second-order theory correctly reproduces the first-order hydrodynamic behavior in the infrared limit while introducing additional nonhydrodynamic relaxation degrees of freedom.

The results of $\mathrm{det}M_{j}=0~(j\neq10)$ in the $k\rightarrow\infty$ limit are given as
\begin{align}
 \omega&=i\left(\frac{1}{4}p-\frac{1}{2}h-\frac{1}{2}\sqrt{\frac{1}{2}p^{2}+\left(q-r\right)+\frac{-p^{3}-4pq+24t}{4h}}\right), \label{eq:43} \\
 \omega&=\pm\frac{1}{3}\sqrt{-\frac{2^{1/3}\tilde{a}_{x}m_{x}^{1/3}+2^{2/3}f_{x}-m_{x}^{2/3}}{2^{1/3}\tau_{16}m_{x}^{1/3}}}k+iY_{x1}, \label{eq:44} \\
 \omega&=\pm\frac{1}{3}\sqrt{-\frac{2^{4/3}m_{x}^{1/3}\tilde{a}_{x}-\left(1+i\sqrt{3}\right)2^{2/3}f_{x}+\left(1-i\sqrt{3}\right)m_{x}^{2/3}}{2^{4/3}m_{x}^{1/3}\tau_{16}}}k+iY_{x2}, \label{eq:45} \\
 \omega&=\pm\frac{1}{3}\sqrt{-\frac{2^{4/3}m_{x}^{1/3}\tilde{a}_{x}-\left(1-i\sqrt{3}\right)2^{2/3}f_{x}+\left(1+i\sqrt{3}\right)m_{x}^{2/3}}{2^{4/3}m_{x}^{1/3}\tau_{16}}}k+iY_{x3}, \label{eq:46} \\
 \omega&=\pm\sqrt{\frac{\gamma_{\perp}}{\tau_{\pi}}}k+\frac{i}{2\tau_{\pi}}, \label{eq:48} \\
 \omega&=i\frac{3\chi_{1}+4\chi_{2}}{3\chi_{1}\tau_{\Sigma_{s}}+4\chi_{2}\tau_{\Sigma}}, \label{eq:49} \\
 \omega&=\pm\sqrt{\frac{-\beta_{(0)}\left(3\tilde{\chi}_{1}\tau_{\Sigma_{s}}+4\tilde{\chi}_{2}\tau_{\Sigma}\right)}{3\tau_{\Sigma}\tau_{\Sigma_{s}}}}k+i\frac{3\tau_{\Sigma_{s}}^{2}\chi_{1}+4\tau_{\Sigma}^{2}\chi_{2}}{6\tau_{\Sigma}\tau_{\Sigma_{s}}^{2}\chi_{1}+8\tau_{\Sigma_{s}}\tau_{\Sigma}^{2}\chi_{2}}, \label{eq:50} \\
 \omega&=i\frac{\chi_{2}+\chi_{3}}{\tau_{\Sigma_{a}}\chi_{2}+\tau_{\Sigma_{s}}\chi_{3}},\label{eq:51} \\
 \omega&=\pm\sqrt{\frac{-\beta_{(0)}\left(\tilde{\chi}_{2}\tau_{\Sigma_{a}}+\tilde{\chi}_{3}\tau_{\Sigma_{s}}\right)}{\tau_{\Sigma_{a}}\tau_{\Sigma_{s}}}}k+i\frac{\tau_{\Sigma_{a}}^{2}\chi_{2}+\tau_{\Sigma_{s}}^{2}\chi_{3}}{2\tau_{\Sigma_{s}}\tau_{\Sigma_{a}}\left(\tau_{\Sigma_{a}}\chi_{2}+\tau_{\Sigma_{s}}\chi_{3}\right)}, \label{eq:52} \\
 \omega&=\pm\sqrt{\frac{\beta_{(0)}\tilde{\chi}_{4}}{\tau_{\Sigma_{4}}}}k+\frac{i}{2\tau_{\Sigma_{4}}} .\label{eq:54}
\end{align}
The parameters in Eq. \eqref{eq:43} are defined in Appendix \ref{chap:5}, and those in Eqs. \eqref{eq:44}-\eqref{eq:46} are defined in Appendix \ref{chap:6} and Appendix \ref{chap:7}. $\mathrm{det}M_{5}=0$ gives Eqs. \eqref{eq:43}-\eqref{eq:46}; $\mathrm{det}M_{6}=0$ gives \eqref{eq:48}; $\mathrm{det}M_{7}=0$ gives Eqs. \eqref{eq:49} and \eqref{eq:50}; $\mathrm{det}M_{8}=0$  gives Eqs. \eqref{eq:51} and \eqref{eq:52}; $\mathrm{det}M_{9}=0$ gives \eqref{eq:54}.

Far more intricate than those in the first-order theory, the dispersion relations \eqref{eq:43}--\eqref{eq:54} reveal a much richer mode structure in the large wave vector limit. In addition to the ordinary soundlike and spin dissipative excitations already present in the zero-spin background theory, the finite spin density generates several additional branches that remain coupled to the background polarization even in the ultraviolet regime. The resulting eigenmodes can be grouped into four categories.

The first category consists of soundlike modes, represented by Eqs. \eqref{eq:48}, \eqref{eq:50}, \eqref{eq:52}, and \eqref{eq:54}, which take the generic form $\omega=\pm vk+i\Gamma$. The corresponding propagation velocities are determined by dissipative transport coefficients and relaxation times, such as $v=\sqrt{\gamma_{\perp}/\tau_{\pi}}$ and $v=\sqrt{-\beta_{(0)}(\tilde{\chi}_{2}\tau_{\Sigma_a}+\tilde{\chi}_{3}\tau_{\Sigma_s})/(\tau_{\Sigma_a}\tau_{\Sigma_s})}$. Since these velocities do not explicitly depend on the spin background parameters, these modes survive unchanged in the limit $S_{(0)}^{\mu\nu}\rightarrow0$ and therefore represent direct extensions of the soundlike excitations already encountered in the zero-spin background theory.

The second category corresponds to spin dissipative modes, represented by Eqs. \eqref{eq:49} and  \eqref{eq:51}. These modes describe purely relaxational dynamics associated with spin transport and spin-current dissipation. Their ultraviolet behavior is governed entirely by the spin dissipative coefficients and the corresponding relaxation times. Consequently, they are insensitive to the background spin density and remain present even in the zero-spin background theory.

The third category contains the complex coupled soundlike modes, represented by Eqs. \eqref{eq:44}-\eqref{eq:46}. Unlike the corresponding modes in the small-$k$ regime, the ultraviolet dispersion relations exhibit a considerably more intricate dependence on the spin background parameters $\mathcal{A}$, $\mathcal{B}$, $\mathcal{C}$, and $\mathcal{D}$. The resulting propagation velocities and damping terms are determined by nontrivial combinations of relaxation times, dissipative transport coefficients, and spin background corrections, indicating that the coupling between hydrodynamic and spin degrees of freedom remains important even at very short wavelengths. 

The fourth category consists of complex coupled spin dissipative modes, represented by Eq. \eqref{eq:43}. In contrast to the analogous modes encountered in the first-order theory, whose large-$k$ behavior is diffusive and therefore acausal, these second-order modes acquire finite relaxation scales through the M$\ddot{\textrm{u}}$ller-Israel-Stewart-type structure of the theory. Their dispersion relations remain explicitly dependent on the spin background parameters, demonstrating that the polarization-induced couplings continue to influence the relaxation dynamics of the spin sector in the ultraviolet limit. Therefore, these modes provide a direct manifestation of the interplay between spin relaxation and causal second-order dissipative dynamics.

Substituting into Eq. \eqref{eq:8} yields the stability condition
\begin{gather}
\tau_{\Pi},\tau_{\pi},\tau_{\Sigma},\tau_{\Sigma_{s}},\tau_{\Sigma_{a}},\tau_{\Sigma_{4}}>0,  \label{eq:55} \\
\gamma_{\perp}>0,\quad \chi_{b}<0, \quad \chi_{s}>0,\\
\frac{1}{2}\gamma_{\parallel}+\frac{\left[\chi_{T}+\theta\left(\beta_{(0)}+\mathcal{C}+\mathcal{D}\right)\right]\left(\tilde{\chi}_{4}\mathcal{B}-\tilde{\chi}_{23}\mathcal{A}\right)}{2c_{v}^{2}}>0,\\
\frac{1}{4}p-\frac{1}{2}h-\frac{1}{2}\sqrt{\frac{1}{2}p^{2}+\left(q-r\right)+\frac{-p^{3}-4pq+24t}{4h}}>0, \label{eq:56} \\
Y_{xi}>0, \label{eq:57}
\end{gather}
and the causality conditions are
\begin{gather}
 0\leq
 c_{c}
 \leq1,\label{eq:102}\\
 0\leq
 -\frac{2^{1/3}\tilde{a}_{x}m_{x}^{1/3}+2^{2/3}f_{x}-m_{x}^{2/3}}{2^{\frac{1}{3}}\tau_{16}m_{x}^{\frac{1}{3}}}
 \leq1, \label{eq:60} \\
 0\leq
 -\frac{2^{4/3}m_{x}^{1/3}\tilde{a}_{x}-\left(1\pm i\sqrt{3}\right)2^{2/3}f_{x}+\left(1\mp i\sqrt{3}\right)m_{x}^{2/3}}{2^{4/3}m_{x}^{1/3}\tau_{16}}
 \leq1, \label{eq:61} \\
 0\leq
 \frac{\gamma_{\perp}}{\tau_{\pi}}
 \leq1, \label{eq:62} \\
 0\leq
 \frac{-\beta_{(0)}\left(3\tilde{\chi}_{1}\tau_{\Sigma_{s}}+4\tilde{\chi}_{2}\tau_{\Sigma}\right)}{3\tau_{\Sigma}\tau_{\Sigma_{s}}}
 \leq1, \label{eq:63} \\
 0\leq
 \frac{-\beta_{(0)}\left(\tilde{\chi}_{2}\tau_{\Sigma_{a}}+\tilde{\chi}_{3}\tau_{\Sigma_{s}}\right)}{\tau_{\Sigma_{a}}\tau_{\Sigma_{s}}}
 \leq1, \label{eq:64} \\
 0\leq
 \frac{\beta_{(0)}\tilde{\chi}_{4}}{\tau_{\Sigma_{4}}}
 \leq1. \label{eq:65}
\end{gather}

 Compared to the acausal outcomes in first-order theory, the structure of the causality conditions in Eqs. \eqref{eq:62}-\eqref{eq:65} reveals that the corresponding causal conditions can be satisfied as long as the relaxation time is sufficiently large.
This is consistent with the intended effect of introducing the relaxation time, and this result has also been reported in previous studies \cite{Sarwar:2022yzs,Xie:2023gbo,Israel:1979wp,10.1098/rspa.1979.0005}. For the conditions in Eqs.\eqref{eq:60} and \eqref{eq:61}, however, however, an intuitive relation for the variation cannot be derived due to the strong coupling between the relaxation time and the spin background state. Nevertheless, from the condition of Eq. \eqref{eq:102}, the following relation can be derived as
\begin{align}
-\frac{\chi_{T}}{\theta}\leq\mathcal{A}+\mathcal{B}\leq\frac{1-\chi_{T}}{\theta},
\end{align}
this gives a constraint condition for the parameters $\mathcal{A}$ and $\mathcal{B}$.

  \subsubsection{Mode analysis along the $z$ direction}
Using the same method, one can derive the characteristic equation along the $z$ axis direction as
\begin{align}
\mathcal{M}_{4}\delta\tilde{X}_{4}=0,
\end{align}
where
\begin{align}
\delta\tilde{X}_{4} \equiv
(&\delta\tilde{e},\delta\tilde{\vartheta}^{z},\delta\tilde{\Sigma},\delta\tilde{\Sigma}_{(s)}^{xx},\delta\tilde{\Sigma}_{(s)}^{yy},\delta\tilde{\Sigma}^{zxy},\delta\tilde{S}^{0z},\delta\tilde{S}^{xy},\delta\tilde{\Pi},\delta\tilde{\pi}^{xx},\delta\tilde{\pi}^{yy}, \notag \\
 &\delta\tilde{\vartheta}^{x},\delta\tilde{\vartheta}^{y},\delta\tilde{\pi}^{xz},\delta\tilde{\pi}^{yz},\delta\tilde{S}^{0x},\delta\tilde{\Sigma}_{(s)}^{xz},\delta\tilde{\Sigma}_{(a)}^{xz}, \notag \\
 &\delta\tilde{S}^{0y},\delta\tilde{\Sigma}_{(s)}^{yz},\delta\tilde{\Sigma}_{(a)}^{yz},\delta\tilde{S}^{xz},\delta\tilde{S}^{yz},\delta\tilde{\Sigma}^{zxz},\delta\tilde{\Sigma}^{zyz}, \notag \\
 &\delta\tilde{\pi}^{xy},\delta\tilde{\Sigma}_{(s)}^{xy},\delta\tilde{\Sigma}_{(a)}^{xy},\delta\tilde{\Sigma}^{xxy},\delta\tilde{\Sigma}^{yxy},\delta\tilde{\Sigma}^{xxz},\delta\tilde{\Sigma}^{yxz},\delta\tilde{\Sigma}^{xyz},\delta\tilde{\Sigma}^{yyz})^T,
\end{align}
and
\begin{align}
 \mathcal{M}_{4}=
 \begin{pmatrix}
 M_{11} & 0 & 0 & 0 & 0 & 0\\
 0 & M_{12} & 0 & 0 & 0 & 0\\
 0 & 0 & M_{13} & 0 & 0 & 0\\
 0 & 0 & 0 & M_{13} & 0 & 0\\
 0 & 0 & 0 & 0 & M_{14} & 0\\
 0 & 0 & 0 & 0 & 0 & M_{15}
 \end{pmatrix},
\end{align}
the explicit expressions for each block matrix in $\mathcal{M}_{4}$ are as follows:
\begin{gather}
M_{11}=
\left(\begin{array}{*{11}{c}}
i\omega&-ik&0&0&0&0&0&0&0&0&0\\-ik\chi_{c}&i\omega&0&0&0&0&-ik\mathcal{A}_{0z}&-ik\mathcal{B}_{xy}&-ik&ik&ik\\-ik\tilde{\chi}_{1}\chi_{T}S_{(0)}^{0z}&0&\Gamma_{3}&0&0&0&-ik\tilde{\chi}_{1}\mathcal{C}_{T}&-ik\tilde{\chi}_{1}S_{(0)}^{0z}\mathcal{D}_{xy}&0&0&0\\-\frac{2}{3}ik\tilde{\chi}_{2}\chi_{T}S_{(0)}^{0z}&0&0&\Gamma_{4}&0&0&-\frac{2}{3}ik\tilde{\chi}_{2}\mathcal{C}_{T}&-\frac{2}{3}ik\tilde{\chi}_{2}S_{(0)}^{0z}\mathcal{D}_{xy}&0&0&0\\-\frac{2}{3}ik\tilde{\chi}_{2}\chi_{T}S_{(0)}^{0z}&0&0&0&\Gamma_{4}&0&-\frac{2}{3}ik\tilde{\chi}_{2}\chi_{b}\mathcal{C}_{T}&-\frac{2}{3}ik\tilde{\chi}_{2}S_{(0)}^{0z}\mathcal{D}_{xy}&0&0&0\\ik\tilde{\chi}_{4}\chi_{T}S_{(0)}^{xy}&0&0&0&0&\Gamma_{6}&ik\tilde{\chi}_{4}S_{(0)}^{xy}\mathcal{C}_{0z}&ik\tilde{\chi}_{4}\mathcal{D}_{T}&0&0&0\\0&-ik\theta S_{(0)}^{0z}&-ik&-ik&-ik&0&i\omega&0&0&0&0\\0&-ik\theta S_{(0)}^{xy}&0&0&0&-ik&0&i\omega&0&0&0\\0&-\left(\gamma_{\parallel}-\frac{4}{3}\gamma_{\perp}\right)ik&0&0&0&0&0&0&-\Gamma_{2}&0&0\\0&-\frac{2}{3}\gamma_{\perp}ik&0&0&0&0&0&0&0&\Gamma_{1}&0\\0&-\frac{2}{3}\gamma_{\perp}ik&0&0&0&0&0&0&0&0&\Gamma_{1}
\end{array}\right)
, \\
M_{12}=
\begin{pmatrix}i\omega & 0 & -ik & 0\\
0 & i\omega & 0 & -ik\\
\gamma_{\perp}ik & 0 & \Gamma_{1} & 0\\
0 & \gamma_{\perp}ik & 0 & \Gamma_{1}
\end{pmatrix}, \\
M_{13}=
\begin{pmatrix}i\omega & ik & -ik\\
ik\tilde{\chi}_{2}\beta_{(0)} & \Gamma_{4} & 0\\
-ik\tilde{\chi}_{3}\beta_{(0)} & 0 & \Gamma_{5}
\end{pmatrix}, \\
M_{14}=
\begin{pmatrix}i\omega & 0 & -ik & 0\\
0 & i\omega & 0 & -ik\\
ik\tilde{\chi}_{4}\beta_{(0)} & 0 & \Gamma_{6} & 0\\
0 & ik\tilde{\chi}_{4}\beta_{(0)} & 0 & \Gamma_{6}
\end{pmatrix}, \\
M_{15}=
\begin{pmatrix}\Gamma_{1} & 0 & 0 & 0 & 0 & 0 & 0 & 0 & 0\\
0 & \Gamma_{4} & 0 & 0 & 0 & 0 & 0 & 0 & 0\\
0 & 0 & \Gamma_{5} & 0 & 0 & 0 & 0 & 0 & 0\\
0 & 0 & 0 & \Gamma_{6} & 0 & 0 & 0 & 0 & 0\\
0 & 0 & 0 & 0 & \Gamma_{6} & 0 & 0 & 0 & 0\\
0 & 0 & 0 & 0 & 0 & \Gamma_{6} & 0 & 0 & 0\\
0 & 0 & 0 & 0 & 0 & 0 & \Gamma_{6} & 0 & 0\\
0 & 0 & 0 & 0 & 0 & 0 & 0 & \Gamma_{6} & 0\\
0 & 0 & 0 & 0 & 0 & 0 & 0 & 0 & \Gamma_{6}
\end{pmatrix}.
\end{gather}

In the $k\rightarrow0$ limit, one obtains
\begin{gather}
 \omega=\frac{i}{\tau_{\pi}} , \label{eq:66} \\
 \omega=\frac{i}{\tau_{\Pi}} , \label{eq:67} \\
 \omega=\frac{i}{\tau_{\Sigma}} , \label{eq:68} \\
 \omega=\frac{i}{\tau_{\Sigma_{s}}} , \label{eq:69} \\
 \omega=\frac{i}{\tau_{\Sigma_{a}}} , \label{eq:70} \\
 \omega=\frac{i}{\tau_{\Sigma_{4}}} , \label{eq:71} \\
 \omega=i\gamma_{\perp}k^{2} , \label{eq:72} \\
 \omega=-i\tilde{\chi}_{23}\beta_{(0)}k^{2} , \label{eq:73} \\
 \omega=i\tilde{\chi}_{4}\beta_{(0)}k^{2} , \label{eq:74} \\
 \omega=\pm c_{v}k+i\left(\frac{1}{2}\gamma_{\parallel}+\frac{\left[\chi_{T}+\theta\left(\beta_{(0)}+\mathcal{C}+\mathcal{D}\right)\right]\left(\tilde{\chi}_{4}\mathcal{B}-\tilde{\chi}_{12}\mathcal{A}\right)}{2c_{v}^{2}}\right)k^{2}, \label{eq:75}
\end{gather}
where Eqs. \eqref{eq:72}-\eqref{eq:75} herein take the same form as the results from the first-order theory, and thus the discrepancies relative to the $x$ direction are also identical. One can obtain the results from $\mathrm{det}M_{j} = 0 ~(j = 11 \sim 14)$ in the $k\rightarrow\infty$ limit as
\begin{align}
 \omega&=i\frac{3\chi_{1}+4\chi_{2}}{3\tau_{\Sigma_{s}}\chi_{1}+4\tau_{\Sigma}\chi_{2}} , \label{eq:76} \\
 \omega&=i\left(\frac{\pm\sqrt{a_{z}^{2}+4b_{z}c_{z}}-a_{z}}{2b_{z}}\right) , \label{eq:77} \\
 \omega&=\pm\frac{1}{3}\sqrt{-\frac{2^{1/3}3\tilde{a}_{z}m_{z}^{1/3}+2^{2/3}f_{z}-m_{z}^{2/3}}{2^{1/3}3\tau_{16}m_{z}^{1/3}}}k+iY_{z1} , \label{eq:78} \\
 \omega&=\pm\frac{1}{3}\sqrt{-\frac{6m_{z}^{1/3}2^{2/3}\tilde{a}_{z}-2\left(1+i\sqrt{3}\right)f_{z}+2^{1/3}\left(1-i\sqrt{3}\right)m_{z}^{2/3}}{6\tau_{16}2^{2/3}m_{z}^{1/3}}}k+iY_{z2} , \label{eq:79} \\
 \omega&=\pm\frac{1}{3}\sqrt{-\frac{6m_{z}^{1/3}2^{2/3}\tilde{a}_{z}-2\left(1-i\sqrt{3}\right)f_{z}+2^{1/3}\left(1+i\sqrt{3}\right)m_{z}^{2/3}}{6\tau_{16}2^{2/3}m_{z}^{1/3}}}k+iY_{z3} , \label{eq:80} \\
 \omega&=\pm\sqrt{\frac{\gamma_{\perp}}{\tau_{\pi}}}k+\frac{i}{2\tau_{\pi}} , \label{eq:82} \\
 \omega&=i\frac{\chi_{2}+\chi_{3}}{\tau_{\Sigma_{a}}\chi_{2}+\tau_{\Sigma_{s}}\chi_{3}} , \label{eq:83} \\
 \omega&=\pm\sqrt{\frac{-\beta_{(0)}\left(\tilde{\chi}_{2}\tau_{\Sigma_{a}}+\tilde{\chi}_{3}\tau_{\Sigma_{s}}\right)}{\tau_{\Sigma_{a}}\tau_{\Sigma_{s}}}}k+i\frac{\tau_{\Sigma_{s}}^{2}\chi_{3}+\tau_{\Sigma_{a}}^{2}\chi_{2}}{2\tau_{\Sigma_{a}}\tau_{\Sigma_{s}}\left(\tau_{\Sigma_{a}}\chi_{2}+\tau_{\Sigma_{s}}\chi_{3}\right)}, \label{eq:84} \\
 \omega&=\pm\sqrt{\frac{\beta_{(0)}\tilde{\chi}_{4}}{\tau_{\Sigma_{4}}}}k+\frac{i}{2\tau_{\Sigma_{4}}}. \label{eq:86}
\end{align}
The parameters in Eq. \eqref{eq:77} are defined in Appendix \ref{chap:8}, while those in Eqs. \eqref{eq:78}–\eqref{eq:80} are defined in Appendixes \ref{chap:9} and \ref{chap:10}. The stability condition is obtained as
\begin{gather}
 \tau_{\Pi},\tau_{\pi},\tau_{\Sigma},\tau_{\Sigma_{s}},\tau_{\Sigma_{a}},\tau_{\Sigma_{4}}>0, \label{eq:87} \\
 \gamma_{\perp}>0,\quad \chi_{b}<0, \quad \chi_{s}>0,\\
 \frac{1}{2}\gamma_{\parallel}+\frac{\left[\chi_{T}+\theta\left(\beta_{(0)}+\mathcal{C}+\mathcal{D}\right)\right]\left(\tilde{\chi}_{4}\mathcal{B}-\tilde{\chi}_{12}\mathcal{A}\right)}{2c_{v}^{2}}>0,\\
 \frac{\pm\sqrt{a_{z}^{2}+4b_{z}c_{z}}-a_{z}}{2b_{z}}>0, \label{eq:88} \\
 Y_{zi}>0. \label{eq:89}
\end{gather}

The causality condition is
\begin{gather}
 0\leq
  c_{c}
  \leq1,\\
 0\leq
 -\frac{2^{\frac{1}{3}}3\tilde{a}_{z}m_{z}^{\frac{1}{3}}+2^{\frac{2}{3}}f_{z}-m_{z}^{\frac{2}{3}}}{2^{\frac{1}{3}}3\tau_{16}m_{z}^{\frac{1}{3}}}
  \leq1, \label{eq:91} \\
 0\leq
 -\frac{6m_{z}^{1/3}2^{2/3}\tilde{a}_{z}-2\left(1\pm i\sqrt{3}\right)f_{z}+2^{1/3}\left(1\mp i\sqrt{3}\right)m_{z}^{2/3}}{6\tau_{16}2^{\frac{2}{3}}m_{z}^{1/3}}
  \leq1, \label{eq:92} \\
 0\leq
 \frac{\gamma_{\perp}}{\tau_{\pi}}
  \leq1, \label{eq:93} \\
 0\leq
 \frac{-\beta_{(0)}\left(\tilde{\chi}_{2}\tau_{\Sigma_{a}}+\tilde{\chi}_{3}\tau_{\Sigma_{s}}\right)}{\tau_{\Sigma_{a}}\tau_{\Sigma_{s}}}
  \leq1, \label{eq:94} \\
 0\leq
 \frac{\beta_{(0)}\tilde{\chi}_{4}}{\tau_{\Sigma_{4}}}
  \leq1. \label{eq:95}
\end{gather}

The physical interpretation of the modes in the $z$ direction largely parallels that discussed for the $x$ direction and will therefore not be repeated here. In the long-wavelength limit, Eqs.~\eqref{eq:72}--\eqref{eq:75} exhibit the same qualitative structure as the corresponding modes obtained in the first-order theory. Consequently, the distinctions between the $x$ and $z$ directions remain identical to those discussed previously. In particular, the spin dissipative modes are governed by different transport coefficients in the two directions, reflecting the anisotropy induced by the nonvanishing spin background.

More significant differences emerge in the large wave-vector limit. Unlike the $x$ direction, where the mode structure contains the additional branches described by Eqs.~\eqref{eq:40} and \eqref{eq:52}, these modes are absent in the $z$ direction. As a consequence, the spectrum of ultraviolet excitations exhibits a clear directional dependence. Furthermore, comparison of Eqs.~\eqref{eq:77}--\eqref{eq:80} with Eqs.~\eqref{eq:43}--\eqref{eq:46} shows that the differences between the two directions cannot be reduced to the simple replacement $\chi_{12}\leftrightarrow\chi_{23}$ that characterizes the first-order theory. Instead, the corresponding dispersion relations possess substantially different structures and depend on distinct combinations of the spin background parameters.

These results indicate that the influence of the finite spin background becomes increasingly pronounced in the second-order theory. Besides modifying the stability and causality conditions through the spin background parameters, the anisotropy associated with the equilibrium spin density leads to genuinely direction-dependent collective excitations. Certain modes appear only for perturbations propagating in specific directions, while different combinations of transport coefficients and spin background corrections govern the dynamics in the transverse and longitudinal sectors.

\section{SUMMARY AND CONCLUSIONS}\label{chap:4}
In this work, we investigated the stability and causality of relativistic spin hydrodynamics in the presence of a nonvanishing spin density background, i.e. $S_{(0)}^{\mu\nu}\neq0$, within the framework of linear mode analysis. Since, under the condition of $\omega^{\mu\nu} \sim \mathcal{O}(\partial)$, the equilibrium state requires the spin chemical potential to be determined by the thermal vorticity $\varpi_{\mu\nu}$, a nonzero spin density would necessarily imply a moving background configuration. In order to maintain a static background in the linear analysis, we instead adopt $\omega^{\mu\nu} \sim \mathcal{O}(1)$, in which case the relation between $\omega^{\mu\nu}$ and $\varpi_{\mu\nu}$ disappears, allowing for a static background with finite spin density. Assuming that the nonvanishing components of the equilibrium spin density are given by  $S_{(0)}^{0z}\neq0$, $S_{(0)}^{xy}\neq0$, the rotational symmetry of the system is reduced from $SO(3)$ to $SO(2)$ in the $x$-$y$ plane. Consequently, separate mode analyses must be performed for perturbations propagating parallel and perpendicular to the spin-polarization direction. From the linearized hydrodynamic equations, the corresponding dispersion relations are derived.

Within first-order spin hydrodynamics, the dispersion relations along the parallel and perpendicular directions exhibit the same structural form in both the small and large wave-vector limits. However, mode \eqref{eq:20} only appears in the $x$ direction. The presence of a finite spin density modifies several dissipative and propagating modes, including their sound velocities and damping coefficients. Different choices of the spin background state can lead to either enhancement or suppression of the damping terms. The difference lies in the transport coefficients entering the dispersion relations: along the $x$ direction they are governed by $\chi_{23}$, whereas along the $z$ direction they involve $\chi_{12}$, indicating that different propagation directions are controlled by different transport coefficients in this anisotropic system. Compared with the vanishing spin density case, entirely new dissipative modes, Eqs. \eqref{eq:24} and \eqref{eq:101}, emerge in addition to these modifications.
However, these modes are acausal, demonstrating that first-order spin hydrodynamics in this framework is intrinsically noncausal.

We further investigated the minimal causal second-order spin hydrodynamics \cite{Xie:2023gbo,Koide:2006ef}, in which only the relaxation-time terms most crucial for causality are retained. For the complete second-order formulation, see Ref. \cite{Biswas:2023qsw}. Our analysis shows that, in the small wave-vector limit, the dispersion relations along both directions resemble those of the first-order theory, with additional contributions arising from the relaxation times. In contrast, in the large wave-vector limit, a finite spin density generates new dissipative and propagating modes, and substantial differences appear between the $x$ and $z$ directions. This demonstrates that the anisotropy induced by the spin density background leads to similar structures at small wave vectors (up to different transport coefficients), while this similarity is broken at large wave vectors.The resulting stability and causality conditions reveal that both the spin density and the spin-related parameters in the thermodynamic equation of state directly determine whether the system is stable and causal.

Finally, our analysis has been restricted to asymptotic limits. The stability and causality properties at finite wave-vector $k$ remain to be explored. As suggested in Ref. \cite{Xie:2023gbo}, systems that are stable and causal in asymptotic regimes may still exhibit instabilities at finite k. In addition, the present study is limited to static backgrounds, and extensions to moving backgrounds \cite{Kovtun:2019hdm,Wang:2023csj} will be addressed in future work.

\begin{acknowledgements}
	 The authors thank Prof. Shi Pu and Dr. Dong-lin Wang for helpful discussions. This work is supported by the National Natural Science Foundation of China (Grants No. 12575144, and No. 11875178).
\end{acknowledgements}

\appendix
\section{Definition of the parameters in Eq. \eqref{eq:43}}\label{chap:5}
\begin{align}
\begin{array}{ccccc}
p=\frac{a_{x}}{f_{1}f_{2}}, & q=\frac{b_{x}}{f_{1}f_{2}}, & r=\frac{c_{x}}{f_{1}f_{2}}, & t=\frac{d_{x}}{f_{1}f_{2}}, & h=\sqrt{\frac{1}{4}p^{2}+\left(q+r\right)},
\end{array}
\end{align}

\begin{align}
  f_{1}=&\left(\beta_{(0)}+\mathcal{C}+\mathcal{D}\right)f_{11}+3\tau_{\pi}\tau_{\Pi}\chi_{T}\left(\mathcal{A}+\mathcal{B}\right), \\
  f_{2}=&\tau_{\pi}\left(\tau_{\Sigma_{a}}\chi_{2}+\tau_{\Sigma_{s}}\chi_{3}\right),
\end{align}
where
\begin{align}
  f_{11}=&4\gamma_{\perp}\left(\tau_{\pi}-\tau_{\Pi}\right)-3\tau_{\pi}\left(\gamma_{\parallel}+\tau_{\Pi}\chi_{c}\right).
\end{align}

\begin{align}
  a_{x}=&\left(\beta_{(0)}+\mathcal{C}+\mathcal{D}\right)a_{x1}+3\tau_{\pi}\chi_{T}\left(\mathcal{A}+\mathcal{B}\right)a_{x2},
\end{align}
where
\begin{align}
  a_{x1}=&4\gamma_{\perp}\left(\tau_{\pi}-\tau_{\Pi}\right)\left[\chi_{2}\left(\tau_{\pi}+\tau_{\Sigma_{a}}\right)+\chi_{3}\left(\tau_{\pi}+\tau_{\Sigma_{s}}\right)\right]-3\tau_{\pi}\left(\chi_{2}a_{xa}+\chi_{3}a_{xs}\right),\\
  a_{xa}=&\gamma_{\parallel}\left(\tau_{\pi}+2\tau_{\Sigma_{a}}\right)+\chi_{c}\left[2\tau_{\Pi}\tau_{\Sigma_{a}}+\tau_{\pi}\left(\tau_{\Pi}+\tau_{\Sigma_{a}}\right)\right], \\
  a_{xs}=&\gamma_{\parallel}\left(\tau_{\pi}+2\tau_{\Sigma_{s}}\right)+\chi_{c}\left[2\tau_{\Pi}\tau_{\Sigma_{s}}+\tau_{\pi}\left(\tau_{\Pi}+\tau_{\Sigma_{s}}\right)\right], \\
  a_{x2}=&2\tau_{\Pi}\left(\tau_{\Sigma_{a}}\chi_{2}+\tau_{\Sigma_{s}}\chi_{3}\right)+\tau_{\pi}\left[(\tau_{\Pi}+\tau_{\Sigma_{a}})\chi_{2}+(\tau_{\Pi}+\tau_{\Sigma_{s}})\chi_{3}\right].
\end{align}

\begin{align}
  b_{x}=&\left(\beta_{(0)}+\mathcal{C}+\mathcal{D}\right)b_{x1}-3\chi_{T}\left(\mathcal{A}+\mathcal{B}\right)\left(\chi_{2}b_{xa}+\chi_{3}b_{xs}\right),
\end{align}
where
\begin{align}
  b_{x1}=&3\chi_{2}\left[\gamma_{\parallel}\left(2\tau_{\pi}+\tau_{\Sigma_{a}}\right)+\chi_{c}b_{xa}\right]+3\chi_{3}\left[\gamma_{\parallel}\left(2\tau_{\pi}+\tau_{\Sigma_{s}}\right)+\chi_{c}b_{xs}\right]-4\gamma_{\perp}\left(\tau_{\pi}-\tau_{\Pi}\right)\left(\chi_{2}+\chi_{3}\right), \\
  b_{xa}=&\tau_{\pi}^{2}+\tau_{\Pi}\tau_{\Sigma_{a}}+2\tau_{\pi}\left(\tau_{\Pi}+\tau_{\Sigma_{a}}\right), \\
  b_{xs}=&\tau_{\pi}^{2}+\tau_{\Pi}\tau_{\Sigma_{s}}+2\tau_{\pi}\left(\tau_{\Pi}+\tau_{\Sigma_{s}}\right).
\end{align}

\begin{align}
  c_{x}=&\left(\beta_{(0)}+\mathcal{C}+\mathcal{D}\right)c_{x1}+3\tau_{\pi}\chi_{T}\left(\mathcal{A}+\mathcal{B}\right)\left(\chi_{2}c_{xa}+\chi_{3}c_{xs}\right),
\end{align}
where
\begin{align}
  c_{x1}=&4\gamma_{\perp}\left(\tau_{\pi}-\tau_{\Pi}\right)\left(\chi_{2}+\chi_{3}\right)-3\tau_{\pi}\left[\chi_{2}\left(\gamma_{\parallel}+\chi_{c}c_{xa}\right)+\chi_{3}\left(\gamma_{\parallel}+\chi_{c}c_{xs}\right)\right], \\
  c_{xa}=&\tau_{\Pi}+\tau_{\Sigma_{a}}, \\
  c_{xs}=&\tau_{\Pi}+\tau_{\Sigma_{s}}.
\end{align}

\begin{align}
  d_{x}=&\left(\beta_{(0)}+\mathcal{C}+\mathcal{D}\right)d_{x1}-\chi_{T}\left(\mathcal{A}+\mathcal{B}\right)\left(\chi_{2}d_{xa}+\chi_{3}d_{xs}\right),
\end{align}
where
\begin{align}
  d_{x1}=&\gamma_{\parallel}\left(\chi_{2}+\chi_{3}\right)+\chi_{c}\left[\chi_{2}\left(2\tau_{\pi}+\tau_{\Pi}+\tau_{\Sigma_{a}}\right)+\chi_{3}\left(2\tau_{\pi}+\tau_{\Pi}+\tau_{\Sigma_{s}}\right)\right], \\
  d_{xa}=&2\tau_{\pi}+\tau_{\Pi}+\tau_{\Sigma_{a}}, \\
  d_{xs}=&2\tau_{\pi}+\tau_{\Pi}+\tau_{\Sigma_{s}}.
\end{align}

\section{Definition of the parameters  at order $k^{1}$ in Eqs. \eqref{eq:44}-\eqref{eq:46}}\label{chap:6}
\begin{align}
  f_{x}=&\left(-\tilde{a}_{x}^{2}+9\tilde{b}_{x}\tau_{16}\right), \\
  m_{x}=&\tilde{c_{x}}+\sqrt{\tilde{c}_{x}^{2}+4f_{x}^{3}},\\
  \tau_{16}=&\tau_{\pi}\tau_{\Pi}\tau_{\Sigma_{s}}\tau_{\Sigma_{a}}\tau_{\Sigma_{4}},
\end{align}
\begin{align}
  \tilde{a}_{x}=&4\gamma_{\perp}\tau_{\Sigma_{s}}\tau_{\Sigma_{a}}\tau_{\Sigma_{4}}(\tau_{\pi}-\tau_{\Pi})+3\tau_{\pi}\{\tau_{\Sigma_{4}}[-\tau_{\Sigma_{s}}\tau_{\Sigma_{a}}(\gamma_{\parallel}+\tau_{\Pi}c^{2}_{v}) \notag \\
  &+\tau_{\Pi}(\tau_{\Sigma_{a}}\tilde{\chi}_{2}+\tau_{\Sigma_{s}}\tilde{\chi}_{3})\mathcal{C}_{T}]-\tau_{\Pi}\tau_{\Sigma_{s}}\tau_{\Sigma_{a}}\tilde{\chi}_{4}\mathcal{D}_{T}\},
\end{align}
\begin{align}
  \tilde{b}_{x}=
  &\tau_{\Sigma_{4}}(\tau_{\Sigma_{a}}\tilde{\chi}_{2}+\tau_{\Sigma_{s}}\tilde{\chi}_{3})\{-4\gamma_{\perp}\tau_{\Pi}\mathcal{C}+\tau_{\pi}[\mathcal{C}(4\gamma_{\perp}-3\gamma_{\parallel}) \notag \\
  &+3\tau_{\Pi}(-\mathcal{C}(\chi_{c}+\theta \mathcal{B})+\mathcal{A}(\chi_{T}+ \theta \mathcal{D}))]\} \notag \\
  &+\tau_{\Sigma_{s}}\tau_{\Sigma_{a}}\tilde{\chi}_{4}[-3\tau_{\pi}\tau_{\Pi}\mathcal{B}(\chi_{T}+ \theta \mathcal{C})+\mathcal{D}\tilde{b}_{x5}] \notag \\
  &-3\beta_{(0)}^{2}\tau_{\pi}\tau_{\Pi}\tilde{\chi}_{4}(\tau_{\Sigma_{a}}\tilde{\chi}_{2}+\tau_{\Sigma_{s}}\tilde{\chi}_{3})+\beta_{(0)}\{\tau_{\Sigma_{4}}(\tau_{\Sigma_{a}}\tilde{\chi}_{2}+\tau_{\Sigma_{s}}\tilde{\chi}_{3})\tilde{b}_{x6} \notag \\
  &+\tilde{\chi}_{4}[\tau_{\Sigma_{s}}\tau_{\Sigma_{a}}\tilde{b}_{x5}-3\tau_{\pi}\tau_{\Pi}(\tau_{\Sigma_{a}}\tilde{\chi}_{2}+\tau_{\Sigma_{s}}\tilde{\chi}_{3})(\mathcal{C}+\mathcal{D})]\},
\end{align}
where
\begin{align}
  \tilde{b}_{x5}=&4\gamma_{\perp}\left(-\tau_{\pi}+\tau_{\Pi}\right)+3\tau_{\pi}\left[\gamma_{\parallel}+\tau_{\Pi}\left(\chi_{c}+\theta \mathcal{A}\right)\right], \\
  \tilde{b}_{x6}=&4\gamma_{\perp}\left(\tau_{\pi}-\tau_{\Pi}\right)-3\tau_{\pi}\left[\gamma_{\parallel}+\tau_{\Pi}\left(\chi_{c}+\theta \mathcal{B}\right)\right].
\end{align}

\begin{align}
 \tilde{c}_{x}=
 &-108\gamma_{\perp}\tau_{\pi}^{2}\tau_{46}\left(\tau_{\pi}-\tau_{\Pi}\right)\left(D_{1}\tau_{\Pi}^{2}+2D_{2}\gamma_{\parallel}\tau_{\Pi}\tau_{46}+2\gamma_{\parallel}^{2}\tau_{46}^{2}\right)+144E\gamma_{\perp}^{2}\tau_{46}^{2}\tau_{\pi}\left(\tau_{\pi}-\tau_{\Pi}\right)^{2}, \notag \\
 &-128\gamma_{\perp}^{3}\tau_{46}^{3}\left(\tau_{\pi}-\tau_{\Pi}\right)^{3}+27\tau_{\pi}^{3}\left[3D_{1}\gamma_{\parallel}\tau_{\Pi}^{2}\tau_{46}+3D_{2}\gamma_{\parallel}^{2}\tau_{\Pi}\tau_{46}^{2}+2\gamma_{\parallel}^{3}\tau_{46}^{3}+\tau_{\Pi}^{3}\left(\tilde{c}_{x1}+\tau_{\Sigma_{s}}^{2}\tilde{c}_{x2}\right)\right],
\end{align}
where
\begin{align}
  \tau_{46}=&\tau_{\Sigma_{s}}\tau_{\Sigma_{a}}\tau_{\Sigma_{4}}, \\
  \tilde{c}_{x1}=
  &-2\mathcal{C}_{T}^{3}\tau_{\Sigma_{a}}^{3}\tau_{\Sigma_{4}}^{3}\tilde{\chi}_{2}^{3}-3\mathcal{C}_{T}\tau_{\Sigma_{s}}\tau_{\Sigma_{a}}^{2}\tau_{\Sigma_{4}}^{2}\tilde{\chi}_{2}^{2}\left(D_{5}\beta_{(0)}\tau_{\Sigma_{a}}+2\mathcal{C}_{T}^{2}\tau_{\Sigma_{4}}\tilde{\chi}_{3}\right) \notag \\
  &+3\beta_{(0)}\tau_{\Sigma_{s}}^{2}\tau_{\Sigma_{a}}\tau_{\Sigma_{4}}\tilde{\chi}_{2}\left(D_{3}\tau_{\Sigma_{a}}^{2}-2D_{4}\beta_{(0)}\tau_{\Sigma_{a}}\tau_{\Sigma_{4}}\tilde{\chi}_{3}-2\beta_{(0)}^{2}\tau_{\Sigma_{4}}^{2}\tilde{\chi}_{3}^{2}\right), \\
  \tilde{c}_{x2}=
  &3D_{3}\beta_{(0)}\tau_{\Sigma_{a}}^{2}\tau_{\Sigma_{4}}\tilde{\chi}_{3}-3D_{5}\beta_{(0)}\mathcal{C}_{T}\tau_{\Sigma_{a}}\tau_{\Sigma_{4}}^{2}\tilde{\chi}_{3}^{2}-2\mathcal{C}_{T}^{3}\tau_{\Sigma_{4}}^{3}\tilde{\chi}_{3}^{3} \notag \\
  &+\tau_{\Sigma_{a}}^{3}\left(c^{2}_{v}\tau_{\Sigma_{4}}+\mathcal{D}_{T}\tilde{\chi}_{4}\right)\left(2c^{4}_{v}\tau_{\Sigma_{4}}^{2}-5\beta_{(0)}\tau_{\Sigma_{4}}\chi_{c}\tilde{\chi}_{4}+2\mathcal{D}_{T}^{2}\tilde{\chi}_{4}^{2}\right), \\
  E=&\tau_{\Sigma_{4}}\left[2\tau_{\Sigma_{s}}\tau_{\Sigma_{a}}\left(\gamma_{\parallel}+c^{2}_{v}\tau_{\Pi}\right)+\mathcal{C}_{T}\tau_{\Pi}(\tau_{\Sigma_{a}}\tilde{\chi}_{2}+\tau_{\Sigma_{s}}\tilde{\chi}_{3})\right]-\mathcal{D}_{T}\tau_{\Pi}\tau_{\Sigma_{s}}\tau_{\Sigma_{a}}\tilde{\chi}_{4},
\end{align}
and
\begin{align}
  D_{1}=
  &\tau_{\Sigma_{4}}^{2}\left(2c^{4}_{v}\tau_{\Sigma_{s}}^{2}\tau_{\Sigma_{a}}^{2}+2\beta_{(0)}\tau_{\Sigma_{s}}\tau_{\Sigma_{a}}(\tau_{\Sigma_{a}}\tilde{\chi}_{2}+\tau_{\Sigma_{s}}\tilde{\chi}_{3})\chi_{c}-\mathcal{C}_{T}^{2}(\tau_{\Sigma_{a}}\tilde{\chi}_{2}+\tau_{\Sigma_{s}}\tilde{\chi}_{3})^{2}\right) \notag \\
  &-2\beta_{(0)}\tau_{46}\tilde{\chi}_{4}\left(\tau_{\Sigma_{s}}\tau_{\Sigma_{a}}\chi_{c}+2\mathcal{C}_{T}(\tau_{\Sigma_{a}}\tilde{\chi}_{2}+\tau_{\Sigma_{s}}\tilde{\chi}_{3})\right)-\mathcal{D}_{T}^{2}\tau_{\Sigma_{s}}^{2}\tau_{\Sigma_{a}}^{2}\tilde{\chi}_{4}^{2}, \\
  D_{2}=&\mathcal{C}_{T}\tau_{\Sigma_{4}}(\tau_{\Sigma_{a}}\tilde{\chi}_{2}+\tau_{\Sigma_{s}}\tilde{\chi}_{3})+\tau_{\Sigma_{s}}\tau_{\Sigma_{a}}\left(2c^{2}_{v}\tau_{\Sigma_{4}}-\mathcal{D}_{T}\tilde{\chi}_{4}\right), \\
  D_{3}=&c^{2}_{v}\tau_{\Sigma_{4}}^{2}\chi_{c}-4\beta_{(0)}\tau_{\Sigma_{4}}\chi_{c}\tilde{\chi}_{4}+\beta_{(0)}^{2}\tilde{\chi}_{4}^{2}, \\
  D_{4}=&\tau_{\Sigma_{4}}\chi_{c}+\beta_{(0)}\tilde{\chi}_{4}, \\
  D_{5}=&\tau_{\Sigma_{4}}\chi_{c}+\mathcal{C}_{T}\tilde{\chi}_{4}.
\end{align}

\section{Definition of the parameters  at order $k^{0}$ in Eqs. \eqref{eq:44}-\eqref{eq:46}}\label{chap:7}
\begin{align}
  Y_{xi}=\frac{Y_{xn}}{Y_{xd}},
\end{align}

\begin{align}
  Y_{xd}=
  &2 \tau _1 (4 \gamma _2 (\tau _1-\tau _2) E_{x15}+3 \tau _1 (-\gamma _3 E_{x15}+\tau _2 (\tau _5 \tilde{\chi}_{2} E_{x16}+\tau _4 (\tau _5 (\tau _6 (5 X^{2}_{xi}-4 \theta  (\mathcal{A}+\mathcal{B})-4 \chi _c) X^{2}_{xi} \notag \\
  &+(-3 \mathcal{B} (\chi_{T}+\theta  \mathcal{C})-E_{x2}E_{x9}) \tilde{\chi}_{4}) X^{2}_{xi}+\tilde{\chi}_{3} E_{x16})))),
\end{align}

\begin{align}
  Y_{xn}=
  &4 \gamma _2 (\tau _1-\tau _2) (\tau _6 (E_{x1} \tau _5 \tilde{\chi}_{2}+\tau _4 (\tau _5 X^{2}_{xi}+E_{x1} \tilde{\chi}_{3})) X^{2}_{xi}-(\tau _5 (E_{x2} \tau _4 X^{2}_{xi}+E_{x7} \tilde{\chi}_{2})+\tau _4 E_{x7} \tilde{\chi}_{3}) \tilde{\chi}_{4}+F_{x8})  \notag \\
  &+3 \tau _1 (\gamma _3 (-F_{x8}-2 (\tau _5 \tilde{\chi}_{2} (X^{2}_{xi} E_{x1} \tau _6-E_{x7} \tilde{\chi}_{4})+\tau _4 (X^{2}_{xi} \tau _6 (\tau _5 X^{2}_{xi}+E_{x1} \tilde{\chi}_{3})-(E_{x2} \tau _5 X^{2}_{xi}+E_{x7} \tilde{\chi}_{3}) \tilde{\chi}_{4})))  \notag \\
  &+2 \tau _2 F_{x7}+\tau _1 (F_{x7}+\tau _2 (\tau _5 (F_{x4}+\tilde{\chi}_{2} F_{x5}) X^{2}_{xi}+\tau _4 (\tau _5 E_{x6} X^{2}_{xi}+\tilde{\chi}_{3} F_{x5}+F_{x4}) X^{2}_{xi}+(\tilde{\chi}_{2}+\tilde{\chi}_{3}) F_{x6}))),
\end{align}

\begin{align}
E_{x1}=&\beta_{(0)}+\mathcal{C},\\
E_{x2}=&\beta_{(0)}+\mathcal{D},\\
E_{x3}=&X^{2}_{xi}-\chi_{c},\\
E_{x4}=&X^{2}_{xi}-\chi_{c}-\theta\mathcal{A},\\
E_{x5}=&X^{2}_{xi}-\chi_{c}-\theta\mathcal{B},\\
E_{x6}=&X^{2}_{xi}-\chi_{c}-\theta\mathcal{A}-\theta\mathcal{B},\\
E_{x7}=&E_{x1}E_{x2}-\mathcal{C}\mathcal{D},\\
E_{x8}=&3X^{2}_{xi}-2\chi_{c},\\
E_{x9}=&4X^{2}_{xi}-3\theta\mathcal{B}-3\chi_{c},\\
E_{x10}=&(X^{2}_{xi}\tau_{6}E_{x9}-(2(\mathcal{A}+\mathcal{B})\chi_{T}+\mathcal{D}E_{x8})\tilde{\chi}_{4})\beta_{0},\\
E_{x11}=&\mathcal{C}(\tau_{6}E_{x9}X^{2}_{xi}+(-2\mathcal{B}\chi_{T}-E_{x2}E_{x8})\tilde{\chi}_{4}),\\
E_{x12}=&S_{1}(3\tau_{6}\chi_{5}(\chi_{T}+\theta\mathcal{D})X^{2}_{xi}+(S_{2}(2\chi_{6}\chi_{T}\chi_{8}+\chi_{9}(2\chi_{5}\chi_{T}+\chi_{8}E_{x8}))-2\mathcal{D}\chi_{5}\chi_{T})\tilde{\chi}_{4}),\\
E_{x13}=&(3X^{2}_{xi}E_{x1}\tau_{6}-2E_{x7}\tilde{\chi}_{4}),\\
E_{x14}=&(4X^{2}_{xi}\tau_{6}-3E_{x2}\tilde{\chi}_{4}),\\
E_{x15}=&(\tau_{5}\tilde{\chi}_{2}E_{x13}+\tau_{4}(\tau_{5}E_{x14}X^{2}_{xi}+\tilde{\chi}_{3}E_{x13})),\\
E_{x16}=&(-E_{x8}\tilde{\chi}_{4}\beta_{0}^{2}+E_{x10}+E_{x11}+E_{x12}),
\end{align}
\begin{align}
F_{x1}=&(X^{2}_{xi}\tau_{6}E_{x5}-((\mathcal{A}+\mathcal{B})\chi_{T}+\mathcal{D}E_{x3})\tilde{\chi}_{4})\beta_{0},\\
F_{x2}=&S_{1}(\tau_{6}\chi_{5}(\chi_{T}+\theta\mathcal{D})X^{2}_{xi}+(S_{2}(\chi_{6}\chi_{T}\chi_{8}+\chi_{9}(\chi_{5}\chi_{T}+\chi_{8}E_{x3}))-\mathcal{D}\chi_{5}\chi_{T})\tilde{\chi}_{4}),\\
F_{x3}=&\mathcal{C}(X^{2}_{xi}\tau_{6}E_{x5}-(\mathcal{B}\chi_{T}+E_{x2}E_{x3})\tilde{\chi}_{4}),\\
F_{x4}=&\tau_{6}E_{x6}X^{2}_{xi}+(-\mathcal{B}(\chi_{T}+\theta\mathcal{C})-E_{x2}E_{x4})\tilde{\chi}_{4},\\
F_{x5}=&(\mathcal{A}(\chi_{T}+\theta\mathcal{D})+\mathcal{C}E_{x5}+\beta_{0}E_{x5}),\\
F_{x6}=&((\chi_{c}-X^{2}_{xi})\tilde{\chi}_{4}\beta_{0}^{2}+F_{x1}+F_{x2}+F_{x3}),\\
F_{x7}=&\tau_{5}\tilde{\chi}_{2}F_{x6}+\tau_{4}(\tau_{5}F_{x4}X^{2}_{xi}+\tilde{\chi}_{3}F_{x6}),\\
F_{x8}=&\tau_{1}(\tau_{5}(\tau_{6}X^{2}_{xi}+E_{x1}\tilde{\chi}_{2}-E_{x2}\tilde{\chi}_{4})X^{2}_{xi}+\tau_{4}((\tau_{5}+\tau_{6})X^{2}_{xi}+E_{x1}\tilde{\chi}_{3}-E_{x2}\tilde{\chi}_{4})X^{2}_{xi} \notag \\
       &+(\tilde{\chi}_{2}+\tilde{\chi}_{3})(X^{2}_{xi}E_{x1}\tau_{6}-E_{x7}\tilde{\chi}_{4})),
\end{align}

\begin{align}
X_{x1}&=\frac{1}{3}\sqrt{-\frac{2^{1/3}\tilde{a}_{x}m_{x}^{1/3}+2^{2/3}f_{x}-m_{x}^{2/3}}{2^{1/3}\tau_{16}m_{x}^{1/3}}},\\
X_{x2}&=\frac{1}{3}\sqrt{-\frac{2^{4/3}m_{x}^{1/3}\tilde{a}_{x}-\left(1+i\sqrt{3}\right)2^{2/3}f_{x}+\left(1-i\sqrt{3}\right)m_{x}^{2/3}}{2^{4/3}m_{x}^{1/3}\tau_{16}}},\\
X_{x3}&=\frac{1}{3}\sqrt{-\frac{2^{4/3}m_{x}^{1/3}\tilde{a}_{x}-\left(1-i\sqrt{3}\right)2^{2/3}f_{x}+\left(1+i\sqrt{3}\right)m_{x}^{2/3}}{2^{4/3}m_{x}^{1/3}\tau_{16}}}.
\end{align}

\section{Definition of the parameters in Eq. \eqref{eq:77}}\label{chap:8}
\begin{align}
a_{z}=
&3\{ \beta_{(0)}[\gamma_{\parallel}+\chi_{c}(\tau_{\pi}+\tau_{\Pi})]+[\gamma_{\parallel}\mathcal{C}-(\tau_{\pi}+\tau_{\Pi})(\mathcal{A}\chi_{T}-\chi_{c}\mathcal{C})] \notag \\
&+[\gamma_{\parallel}\mathcal{D}-(\tau_{\pi}+\tau_{\Pi})(\mathcal{B}\chi_{T}-\chi_{c}\mathcal{D})]\}, \\
b_{z}=&\left[\beta_{(0)}b_{z1}+\left(3\tau_{\pi}\tau_{\Pi}\mathcal{A}\chi_{T}+\mathcal{C}b_{z1}\right)+\left(3\tau_{\pi}\tau_{\Pi}\mathcal{B}\chi_{T}+\mathcal{D}b_{z1}\right)\right], \\
b_{z1}=&4\gamma_{\perp}\left(\tau_{\pi}-\tau_{\Pi}\right)-3\tau_{\pi}\left(\gamma_{\parallel}+\tau_{\Pi}\chi_{c}\right), \\
c_{z}=&3\left[\chi_{c}\left(\beta_{(0)}+\mathcal{C}+\mathcal{D}\right)-\chi_{T}\left(\mathcal{A}+\mathcal{B}\right)\right].
\end{align}

\section{Definition of the parameters  at order $k^{1}$ in Eqs. \eqref{eq:78}-\eqref{eq:80}}\label{chap:9}
\begin{align}
f_{z}=&-9\tilde{a_{z}}^{2}+27\tilde{b}_{z}\tau_{16}, \\
m_{z}=&\tilde{c}_{z}+\sqrt{\tilde{c}_{z}^{2}+4f_{z}^{3}},
\end{align}

\begin{align}
\tilde{a}_{z}=
&4\gamma_{\perp}\tau_{\Sigma_{s}}\tau_{\Sigma}\tau_{\Sigma_{4}}(\tau_{\pi}-\tau_{\Pi})+3\tau_{\pi}\{\tau_{\Sigma_{4}}[-\tau_{\Sigma_{s}}\tau_{\Sigma}(\gamma_{\parallel}+\tau_{\Pi}c^{2}_{v}) \notag \\
&+\tau_{\Pi}(\tau_{\Sigma_{s}}\tilde{\chi}_{1}+\frac{4}{3}\tau_{\Sigma}\tilde{\chi}_{2})(\beta_{(0)}+\mathcal{C})]-\tau_{\Pi}\tau_{\Sigma_{s}}\tau_{\Sigma}\tilde{\chi}_{4}(\beta_{(0)}+\mathcal{D})\},
\end{align}

\begin{align}
\tilde{b}_{z}=
&\tau_{\Sigma_{4}}\tilde{b}_{z4}\{-4\gamma_{\perp}\tau_{\Pi}\mathcal{C}+\tau_{\pi}[\mathcal{C}(4\gamma_{\perp}-3\gamma_{\parallel})+3\tau_{\Pi}(-\mathcal{C}(\chi_{c}+\theta \mathcal{B})+\mathcal{A}(\chi_{T}+\theta \mathcal{D}))]\} \notag \\
&+3\tau_{\Sigma_{s}}\tau_{\Sigma}\tilde{\chi}_{4}[-3\tau_{\pi}\tau_{\Pi}\mathcal{B}(\chi_{T}+\theta \mathcal{C})+\mathcal{D}\tilde{b}_{z5}]-3\beta_{(0)}^{2}\tau_{\pi}\tau_{\Pi}\tilde{\chi}_{4}\tilde{b}_{z4} \notag \\
&+\beta_{(0)}\{\tau_{\Sigma_{4}}\tilde{b}_{z4}\tilde{b}_{z6}+3\tilde{\chi}_{4}[\tau_{\Sigma_{s}}\tau_{\Sigma}\tilde{b}_{z5}-\tau_{\pi}\tau_{\Pi}\tilde{b}_{z4}(\mathcal{C}+\mathcal{D})]\}, \\
\tilde{b}_{z4}=&(3\tau_{\Sigma_{s}}\tilde{\chi}_{1}+4\tau_{\Sigma}\tilde{\chi}_{2}),\\
\tilde{b}_{z5}=&4\gamma_{\perp}(-\tau_{\pi}+\tau_{\Pi})+3\tau_{\pi}[\gamma_{\parallel}+\tau_{\Pi}(\chi_{c}+\theta \mathcal{A})], \\
\tilde{b}_{z6}=&4\gamma_{\perp}(\tau_{\pi}-\tau_{\Pi})-3\tau_{\pi}[\gamma_{\parallel}+\tau_{\Pi}(\chi_{c}+\theta \mathcal{B})].
\end{align}

\begin{align}
\tilde{c}_{z}=
&-324\gamma_{\perp}\tau_{\pi}^{2}\tau_{36}(\tau_{\pi}-\tau_{\Pi})(-D_{z}\tau_{\Pi}^{2}+6B_{z}\gamma_{\parallel}\tau_{\Pi}\tau_{36}+18\gamma_{\parallel}^{2}\tau_{36}^{2})+1296\gamma_{\perp}^{2}\tau_{\pi}(\tau_{\pi}-\tau_{\Pi})^{2}\tau_{36}^{2}(\tau_{\Pi}\tilde{B}_{z}+6\gamma_{\parallel}\tau_{36}) \notag \\
&-3456\gamma_{\perp}^{3}(\tau_{\pi}-\tau_{\Pi})^{3}\tau_{36}^{3}-27\tau_{\pi}^{3}[9\gamma_{\parallel}\tau_{\Pi}^{2}\tau_{36}\tilde{D}_{z}-27\gamma_{\parallel}^{2}\tau_{\Pi}\tau_{36}^{2}\tilde{B}_{z}-54\gamma_{\parallel}^{3}\tau_{36}^{3}+\tau_{\Pi}^{3}(\tilde{c}_{z1}+\tilde{c}_{z2}+\tilde{c}_{z3}+\tilde{c}_{z4}) \notag \\
&+9\tau_{\Sigma}^{2}\tau_{\Sigma_{s}}\tau_{\Sigma_{4}}\tilde{\chi}_{1}(\beta_{(0)}^{3}\tilde{c}_{z5}+3\beta_{(0)}^{2}\tilde{c}_{z6}+\tilde{c}_{z7}-3\beta_{(0)}\tilde{c}_{z8})], \\
\tilde{c}_{z1}=&-128\mathcal{C}_{T}^{3}\tau_{\Sigma_{4}}^{3}\tilde{\chi}_{2}^{3}, \\
\tilde{c}_{z2}=&-144\mathcal{C}_{T}\tau_{\Sigma_{s}}\tau_{\Sigma_{4}}^{2}\tilde{\chi}_{2}^{2}(\tilde{e}_{z1}\tau_{\Sigma_{4}}+\tilde{\chi}_{4}(\beta_{(0)}\mathcal{C}_{T}+\mathcal{D}(\beta_{(0)}-2\mathcal{C}))), \\
\tilde{c}_{z3}=
&27\tau_{\Sigma_{s}}^{3}(c^{2}_{v}\tau_{\Sigma_{4}}+\mathcal{D}_{T}\tilde{\chi}_{4})(2c^{4}_{v}\tau_{\Sigma_{4}}^{2}+2\mathcal{D}_{T}^{2}\tilde{\chi}_{4}^{2}+\tau_{\Sigma_{4}}\tilde{\chi}_{4}(\beta_{(0)}(4\theta \mathcal{B}-5(\theta \mathcal{A}+\chi_{c})) \notag \\
&+(9\mathcal{B}(\theta \mathcal{C}+\chi_{T})+\mathcal{D}(4\theta \mathcal{B}-5(\theta \mathcal{A}+\chi_{c}))))), \\
\tilde{c}_{z4}=&-108\tau_{\Sigma_{s}}^{2}\tau_{\Sigma_{4}}\tilde{\chi}_{2}(-\beta_{(0)}^{3}\tilde{\chi}_{4}^{2}-\beta_{(0)}^{2}\tilde{\chi}_{4}\tilde{c}_{z41}+\tilde{c}_{z42}+\beta_{(0)}\tilde{c}_{z43}),
\end{align}

\begin{align}
\tilde{c}_{z41}=&2\tau_{\Sigma_{4}}(\theta\mathcal{A}+\theta\mathcal{B}-2\chi_{c})+\tilde{\chi}_{4}(\mathcal{C}+2\mathcal{D}), \\
\tilde{c}_{z42}=
&-c^{2}_{v}\tau_{\Sigma_{4}}^{2}(\mathcal{C}(\theta\mathcal{B}+\chi_{c})-\mathcal{A}(2\theta\mathcal{C}+3\theta\mathcal{D}+3\chi_{T}))+\tau_{\Sigma_{4}}\tilde{\chi}_{4}(\mathcal{B}\mathcal{C}(3\theta\mathcal{C}+\theta\mathcal{D}+3\chi_{T}) \notag \\
&+\mathcal{D}(\mathcal{A}(\theta\mathcal{C}+3\theta\mathcal{D}+3\chi_{T})-2\chi_{c}\mathcal{C}))+2\mathcal{C}\mathcal{D}^{2}\tilde{\chi}_{4}^{2}, \\
\tilde{c}_{z43}=
&-c^{2}_{v}\tau_{\Sigma_{4}}^{2}(-2\theta\mathcal{A}+\theta\mathcal{B}+\chi_{c})+\mathcal{D}\tilde{\chi}_{4}^{2}(\mathcal{C}-\mathcal{D})-\tau_{\Sigma_{4}}\tilde{\chi}_{4}(2\mathcal{D}(\theta\mathcal{A}-2\chi_{c}) \notag \\
&+\mathcal{B}(-3\theta\mathcal{C}+2\theta\mathcal{D}+6\chi_{T})+2\mathcal{C}(\theta\mathcal{B}-2\chi_{c})+\mathcal{A}(2\theta\mathcal{C}-3\theta\mathcal{D}+6\chi_{T})),
\end{align}

\begin{align}
\tilde{c}_{z5}=&32\tau_{\Sigma_{4}}^{2}\tilde{\chi}_{2}^{2}+24\tau_{\Sigma_{s}}\tau_{\Sigma_{4}}\tilde{\chi}_{4}\tilde{\chi}_{2}-9\tau_{\Sigma_{s}}^{2}\tilde{\chi}_{4}^{2}, \\
\tilde{c}_{z6}=
&-3\tau_{\Sigma_{s}}^{2}\tilde{\chi}_{4}(2\tau_{\Sigma_{4}}(\theta\mathcal{A}+\theta\mathcal{B}-2\chi_{c})+\tilde{\chi}_{4}(\mathcal{C}+2\mathcal{D})) \notag \\
&+8\tau_{\Sigma_{4}}\tau_{\Sigma_{s}}\tilde{\chi}_{2}(\tau_{\Sigma_{4}}(-2\theta\mathcal{A}+\theta\mathcal{B}+\chi_{c})+\tilde{\chi}_{4}(2\mathcal{C}+\mathcal{D}))+32\mathcal{C}\tau_{\Sigma_{4}}^{2}\tilde{\chi}_{2}^{2}, \\
\tilde{c}_{z7}=
&9\tau_{\Sigma_{s}}^{2}(-c^{2}_{v}\tau_{\Sigma_{4}}^{2}(\mathcal{C}(\theta\mathcal{B}+\chi_{c})-\mathcal{A}(2\theta\mathcal{C}+3\theta\mathcal{D}+3\chi_{T}))+\tau_{\Sigma_{4}}\tilde{\chi}_{4}(\mathcal{B}\mathcal{C}(3\theta\mathcal{C}+\theta\mathcal{D}+3\chi_{T}) \notag \\
&+\mathcal{D}(\mathcal{A}(\theta\mathcal{C}+3\theta\mathcal{D}+3\chi_{T})-2\chi_{c}\mathcal{C}))+2\mathcal{D}^{2}\mathcal{C}\tilde{\chi}_{4}^{2}) \notag \\
&+32\mathcal{C}^{3}\tau_{\Sigma_{4}}^{2}\tilde{\chi}_{2}^{2}-24\mathcal{C}\tau_{\Sigma_{s}}\tau_{\Sigma_{4}}\tilde{\chi}_{2}(\tau_{\Sigma_{4}}(\mathcal{A}(2\theta\mathcal{C}+3\theta\mathcal{D}+3\chi_{T})-\mathcal{C}(\theta\mathcal{B}+\chi_{c}))+2\mathcal{C}\mathcal{D}\tilde{\chi}_{4}), \\
\tilde{c}_{z8}=
&3\tau_{\Sigma_{s}}^{2}(c^{2}_{v}\tau_{\Sigma_{4}}^{2}(-2\theta\mathcal{A}+\theta\mathcal{B}+\chi_{c})+\tau_{\Sigma_{4}}\tilde{\chi}_{4}(2\mathcal{D}(\theta\mathcal{A}-2\chi_{c}) \notag \\
&+\mathcal{B}(-3\theta\mathcal{C}+2\theta\mathcal{D}+6\chi_{T})+2\mathcal{C}(\theta\mathcal{B}-2\chi_{c}) \notag \\
&+\mathcal{A}(2\theta\mathcal{C}-3\theta\mathcal{D}+6\chi_{T}))+\mathcal{D}\chi_{4}^{2}\chi_{s}^{2}(\mathcal{D}-\mathcal{C})) \notag \\
&+8\tau_{\Sigma_{4}}\tau_{\Sigma_{s}}\tilde{\chi}_{2}(\tau_{\Sigma_{4}}(\mathcal{A}(4\theta\mathcal{C}+3\theta\mathcal{D}+3\chi_{T})-2\mathcal{C}(\theta\mathcal{B}+\chi_{c})) \notag \\
&+\chi_{4}\mathcal{C}\chi_{s}(\mathcal{D}-\mathcal{C}))-32\mathcal{C}^{2}\tau_{\Sigma_{4}}^{2}\tilde{\chi}_{2}^{2},
\end{align}

\begin{align}
\tilde{B}_{z}=&\tau_{\Sigma_{4}}(6c^{2}_{v}\tau_{\Sigma}\tau_{\Sigma_{s}}+\mathcal{C}_{T}\chi_{b}\tau_{z})-3\mathcal{D}_{T}\tau_{\Sigma}\tau_{\Sigma_{s}}\tilde{\chi}_{4},
\end{align}

\begin{align}
\tilde{D}_{z}=
&9\mathcal{C}_{T}^{2}\tau_{\Sigma_{s}}^{2}\tau_{\Sigma_{4}}^{2}\tilde{\chi}_{1}^{2}+3\tau_{36}\tilde{\chi}_{1}(\tau_{\Sigma_{4}}(8\mathcal{C}_{T}^{2}\tilde{\chi}_{2}+3\tau_{\Sigma_{s}}\tilde{D}_{z1})+6\tau_{\Sigma_{s}}\tilde{\chi}_{4}\tilde{D}_{z2}) \notag \\
&+\tau_{\Sigma}^{2}(16\mathcal{C}_{T}^{2}\tau_{\Sigma_{4}}^{2}\tilde{\chi}_{2}^{2}+12\tau_{\Sigma_{s}}\tau_{\Sigma_{4}}\tilde{\chi}_{2}(\tau_{\Sigma_{4}}\tilde{D}_{z1}+2\tilde{\chi}_{4}\tilde{D}_{z2})+9\tau_{\Sigma_{s}}^{2}\tilde{D}_{z3}), \\
\tilde{D}_{\text{z1}}=
&\beta_{(0)}(\theta\mathcal{A}-2\theta\mathcal{B}-2\chi_{c})+\mathcal{A}(\theta\mathcal{C}+3\theta\mathcal{D}+3\chi_{T})-2\mathcal{C}(\theta\mathcal{B}+\chi_{c}), \\
\tilde{D}_{\text{z2}}=&2\beta_{(0)}\mathcal{C}_{T}+\mathcal{D}(2\beta_{(0)}-\mathcal{C}), \\
\tilde{D}_{\text{z3}}=
&-2c^{4}_{v}\tau_{\Sigma_{4}}^{2}+\mathcal{D}_{T}^{2}\tilde{\chi}_{4}^{2}+\tau_{\Sigma_{4}}\tilde{\chi}_{4}(\beta_{(0)}(2\theta\mathcal{A}-\theta\mathcal{B}+2\chi_{c}) \notag \\
&+2\mathcal{D}(\theta\mathcal{A}+\chi_{c})-\mathcal{B}(3\theta\mathcal{C}+\theta\mathcal{D}+3\chi_{T})),
\end{align}

\begin{align}
\tilde{e}_{z1}=
&\beta_{(0)}(-2\theta\mathcal{A}+\theta\mathcal{B}+\chi_{c})+\mathcal{C}(\theta\mathcal{B}+\chi_{c})-\mathcal{A}(2\theta\mathcal{C}+3\theta\mathcal{D}+3\chi_{T}),
\end{align}

\begin{align}
\tau_{36}=&\tau_{\Sigma}\tau_{\Sigma_{s}}\tau_{\Sigma_{4}}, \\
\tau_{z}=&3\tau_{\Sigma_{s}}\chi_{1}+4\tau_{\Sigma}\chi_{2}.
\end{align}

\section{Definition of the parameters  at order $k^{0}$ in Eqs. \eqref{eq:78}-\eqref{eq:80}}\label{chap:10}
\begin{align}
  Y_{zi}=\frac{Y_{zn}}{Y_{zd}},
\end{align}

\begin{align}
  Y_{zd}=
  &\tau_{\pi} \tau_{\Sigma_{s}} (4 \gamma_{\perp} (\tau_{\pi}-\tau_{\Pi}) (G_{z2}-G_{z3})+3 \tau_{\pi} (-\gamma_{\parallel} (G_{z2}-G_{z3})+\tau_{\Pi} (3 \tau_{\Sigma_{s}} \tilde{\chi}_{1} G_{z6} \notag \\
  &+\tau_{\Sigma} (3 X_{zi}^{2} \tau_{\Sigma_{s}} (\tilde{\chi}_{4} (-H_{z2} G_{z4}-7 \mathcal{B} (\theta \mathcal{C}+\chi_{T}))+X_{zi}^{2} \tau_{\Sigma_{4}} (2 X_{zi}^{2}+H_{z7}))+4 \tilde{\chi}_{2} G_{z6})))),
\end{align}
\begin{align}
  Y_{zn}=
  & (4 \gamma_{\perp} (\tau_{\pi}-\tau_{\Pi}) (\tau_{\Sigma_{s}} H_{z12}+H_{z16})+3 \tau_{\pi} (-\gamma_{\parallel} (2 \tau_{\Sigma_{s}} H_{z12}+H_{z16})+2 \tau_{\Pi} \tau_{\Sigma_{s}} H_{z11} \notag \\
  &+\tau_{\pi} (\tau_{\Sigma_{s}} H_{z11}+\tau_{\Pi} (\tau_{\Sigma_{s}} (H_{z17}+2 (3 \tilde{\chi}_{1}+2 \tilde{\chi}_{2})  H_{z8})+\tau_{\Sigma} (3 \tau_{\Sigma_{s}}^2 H_{z7} X_{zi}^{4}+H_{z18}+4\tilde{\chi}_{2} H_{z8}))))),
\end{align}

\begin{align}
G_{z1}=&3\tau_{\Sigma_{s}}\chi_{1}+4\tau_{\Sigma}\chi_{2},\\
G_{z2}=&X_{zi}^{2}\tau_{\Sigma_{4}}(27X_{zi}^{2}\tau_{\Sigma}\tau_{\Sigma_{s}}+7\chi_{b}G_{z1}H_{z1}),\\
G_{z3}=&\tilde{\chi}_{4}(21X_{zi}^{2}\tau_{\Sigma}\tau_{\Sigma_{s}}H_{z2}+5\beta_{(0)}\chi_{b}G_{z1}H_{z3}),\\
G_{z4}=&9X_{zi}^{2}-7\theta\mathcal{A}-7\chi_{c},\\
G_{z5}=&9X_{zi}^{2}-7\theta\mathcal{B}-7\chi_{c},\\
G_{z6}=&(X_{zi}^{2}\tau_{\Sigma_{4}}(G_{z5}H_{z1}+7\mathcal{A}\chi_{T}+7\theta\mathcal{A}\mathcal{D})-\beta_{(0)}\tilde{\chi}_{4}((7X_{zi}^{2}-5\chi_{c})H_{z3}+5\chi_{T}(\mathcal{A}+\mathcal{B}))),
\end{align}
\begin{align}
H_{z1}=&\beta_{(0)}+\mathcal{C},\\
H_{z2}=&\beta_{(0)}+\mathcal{D},\\
H_{z3}=&\beta_{(0)}+\mathcal{C}+\mathcal{D},\\
H_{z4}=&X_{zi}^{2}-\chi_{c},\\
H_{z5}=&X_{zi}^{2}-\chi_{c}-\theta\mathcal{A},\\
H_{z6}=&X_{zi}^{2}-\chi_{c}-\theta\mathcal{B},\\
H_{z7}=&X_{zi}^{2}-\chi_{c}-\theta\mathcal{A}-\theta\mathcal{B},\\
H_{z8}=&X_{zi}^{2}\tau_{\Sigma_{4}}H_{z9}-\beta_{(0)}\chi_{4}\chi_{s}((\mathcal{A}+\mathcal{B})\chi_{T}+H_{z4}H_{z3}),\\
H_{z9}=&\mathcal{A}\chi_{T}+H_{z6}H_{z1}+\theta\mathcal{A}\mathcal{D},\\
H_{z10}=&\beta_{(0)}H_{z5}+H_{z5}\mathcal{D}+\mathcal{B}(\chi_{T}+\theta\mathcal{C}),\\
H_{z11}=&3\tau_{\Sigma_{s}}\tilde{\chi}_{1}H_{z8}+\tau_{\Sigma}(3\tau_{\Sigma_{s}}(\tau_{\Sigma_{4}}H_{z7}X_{zi}^{2}-\tilde{\chi}_{4}H_{z10})X_{zi}^{2}+4\tilde{\chi}_{2}H_{z8}),\\
H_{z12}=&X_{zi}^{2}\tau_{\Sigma_{4}}(3\tau_{\Sigma}\tau_{\Sigma_{s}}X_{zi}^{2}+G_{z1}H_{z1}\chi_{b})-\tilde{\chi}_{4}(3\tau_{\Sigma}\tau_{\Sigma_{s}}H_{z2}X_{zi}^{2}+\beta_{(0)}G_{z1}H_{z3}\chi_{b}),\\
H_{z13}=&3\tau_{\Sigma_{s}}c_{v}^{4}(\tau_{\Sigma_{s}}\tau_{\Sigma_{4}}+\tau_{\Sigma}(\tau_{\Sigma_{s}}+2\tau_{\Sigma_{4}})),\\
H_{z14}=&H_{z1}\chi_{b}X_{zi}^{2}(3\tau_{\Sigma_{s}}\chi_{1}(\tau_{\Sigma_{s}}+2\tau_{\Sigma_{4}})+4\chi_{2}(\tau_{\Sigma_{s}}\tau_{\Sigma_{4}}+\tau_{\Sigma}(\tau_{\Sigma_{s}}+\tau_{\Sigma_{4}}))),\\
H_{z15}=&\tilde{\chi}_{4}(3\tau_{\Sigma_{s}}(2\tau_{\Sigma}+\tau_{\Sigma_{s}})H_{z2}X_{zi}^{2}+2\beta_{(0)}H_{z3}\chi_{b}(3\tau_{\Sigma_{s}}\chi_{1}+2\chi_{2}(\tau_{\Sigma}+\tau_{\Sigma_{s}}))),\\
H_{z16}=&\tau_{\pi}(H_{z13}+H_{z14}-H_{z15}),\\
H_{z17}=&3\tau_{\Sigma_{s}}X_{zi}^{2}(\tau_{\Sigma_{4}}H_{z7}X_{zi}^{2}+\tilde{\chi}_{1}H_{z9}-\tilde{\chi}_{4}H_{z10}),\\
H_{z18}=&2\tau_{\Sigma_{s}}X_{zi}^{2}(3\tau_{\Sigma_{4}}H_{z7}X_{zi}^{2}+2\tilde{\chi}_{2}H_{z9}-3\tilde{\chi}_{4}H_{z10}),
\end{align}
\begin{align}
  X_{z1}&=\frac{1}{3}\sqrt{-\frac{2^{\frac{1}{3}}3\tilde{a}_{z}m_{z}^{\frac{1}{3}}+2^{\frac{2}{3}}f_{z}-m_{z}^{\frac{2}{3}}}{2^{\frac{1}{3}}3\tau_{16}m_{z}^{\frac{1}{3}}}},\\
  X_{z2}&=\frac{1}{3}\sqrt{-\frac{6m_{z}^{\frac{1}{3}}2^{\frac{2}{3}}\tilde{a}_{z}-2\left(1+i\sqrt{3}\right)f_{z}+2^{\frac{1}{3}}\left(1-i\sqrt{3}\right)m_{z}^{\frac{2}{3}}}{6\tau_{16}2^{\frac{2}{3}}m_{z}^{\frac{1}{3}}}},\\
  X_{z3}&=\frac{1}{3}\sqrt{-\frac{6m_{z}^{\frac{1}{3}}2^{\frac{2}{3}}\tilde{a}_{z}-2\left(1-i\sqrt{3}\right)f_{z}+2^{\frac{1}{3}}\left(1+i\sqrt{3}\right)m_{z}^{\frac{2}{3}}}{6\tau_{16}2^{\frac{2}{3}}m_{z}^{\frac{1}{3}}}}.
\end{align}

\bibliography{References}

\end{document}